# HFODD

## (v2.08k)

## User's Guide

# HFODD (v2.08k): User's Guide.

J. Dobaczewski[a,b,c,d,1], J. Dudek[b,2], and P. Olbratowski[a,b,3]

[a]*Institute of Theoretical Physics, Warsaw University*
*ul. Hoża 69, PL-00681 Warsaw, Poland*
[b]*Institut de Recherches Subatomiques, CNRS-IN$_2$P$_3$/Université Louis Pasteur,*
*F-67037 Strasbourg Cedex 2, France*
[c]*Department of Physics and Astronomy, The University of Tennessee,*
*Knoxville, Tennessee 37996, USA*
[d]*Physics Division, Oak Ridge National Laboratory,*
*P.O. Box 2008, Oak Ridge, Tennessee 37831, USA*

## Abstract

We describe the input data and installation procedures of the code HFODD (v2.08k). The present write-up contains complete and comprehensive information that has originally been given in five independent publications. It is enhanced by the subject index and indexes of variables, input-data keywords, subroutines, and files that are used in this user guide.

## Keywords

Hartree-Fock; Hartree-Fock-Bogolyubov; Skyrme interaction; self-consistent mean field; nuclear many-body problem; superdeformation; quadrupole deformation; octupole deformation; pairing; nuclear radii; single-particle spectra; nuclear rotation; high-spin states; three-dimensional rotation; chiral symmetry in nuclei; gauge symmetry; moments of inertia; level crossings; harmonic oscillator; Coulomb field; pairing; point-group symmetries.

---

[1]E-mail: jacek.dobaczewski@fuw.edu.pl
[2]E-mail: jerzy.dudek@ires.in2p3.fr
[3]E-mail: przemyslaw.olbratowski@fuw.edu.pl

# Contents

# 1.   INTRODUCTION

The code HFODD (v2.08k) solves the nuclear Skyrme-Hartree-Fock or Skyrme-Hartree-Fock-Bogolyubov problem by using the Cartesian deformed harmonic-oscillator basis. It is a result of a long-term project that is developed at the Institute of Theoretical Physics, Warsaw University and Institut de Recherches Subatomiques, Université Louis Pasteur in Strasbourg. The code has been published and is available from the Computer Physics Communications Program Library (http://www.cpc.cs.qub.ac.uk/cpc/). The description of the code and how to use it have been published in the Computer Physics Communications in the form of five independent publications, which are below referred to as I [1], II [2], III [3], IV [4], and V [5].

The present user's guide summarizes articles I-IV and provides practical details presented in the form that is suitable as a rapid reference on how to run the code. Information pertaining to the physics issues and numerical methods is not repeated here, and the user is referred to the original publications and to the references given therein.

Together with the publication of the present user's guide we have established the HFODD home page (http://www.fuw.edu.pl/~dobaczew/hfodd/hfodd.html), where more information is available, and which will contain information on future versions and extensions being developed.

# 2. INPUT DATA FILE

The code reads the input data from the standard FORTRAN input file, and a general structure of the input data file is defined by the following rules:

1. Input data file is an ASCII file composed of independent input items.

2. An item is composed of the keyword line, and of the data line which follows immediately the keyword line. Two items contain no data lines, see point 6 below.

3. The keyword line contains in columns 1 through 10 a keyword which is a specific text defining the item. If the keyword has less than 10 characters it has to be obligatorily padded with appropriate number of trailing spaces. Columns beyond 10 are ignored and can be used to place arbitrary comments or texts.

4. The data line contains a prescribed number of data values which are read in the free FORTRAN FORMAT. The type of data (`REAL`, `INTEGER`, or `CHARACTER`) should match the definition of the given data line. The `CHARACTER` data should be provided starting from the 13-th column of the data line.

5. Items can be separated by an arbitrary numbers of comment lines. A line is treated as a comment and ignored provided it does not contain in columns 1 through 10 any of the valid keywords.

6. Two items contain no data lines. The first one is defined by the keyword `EXECUTE`, and starts a calculation for the currently defined input parameters. The second one is defined by the keyword `ALL_DONE`, and terminates the program. The latter item is not required if the compiler is able to properly recognize the end of input data file. When the end of input data file is encountered, the program terminates as if the keyword `ALL_DONE` was found.

7. A given item may occur any number of times in different places of the input data file. Only the last one of the same items occurring before a given `EXECUTE` keyword is taken into account.

8. The items appearing between two consecutive items `EXECUTE` define the data set, i.e, the set of input parameters, for which the calculation will start at the moment when the latter of the two such items is found. Between two consecutive items `EXECUTE`, the order of other items is arbitrary.

9. All input parameters have the default values defined in the code (subroutine `PREDEF`). Therefore, if the input data file contains only the item `EXECUTE`, and no other lines, a calculation for the default values is performed. In the present version, this results in performing 50 iterations for the superdeformed state of $^{152}$Dy at the angular frequency of $\hbar\omega$=0.5 MeV.

10. Only those items which define the values of input parameters which are different from the default values have to be included in the input data file. The values of input parameters defined by any item stay in effect till another occurrence of the same item, or till the code terminates. This rule is valid irrespective of how many EXECUTE items follow the given item.

In the following subsections we list all the valid keywords and their corresponding data lines. Titles of the subsections are introduced only to organize the keywords into groups pertaining to specific categories, but are otherwise irrelevant. The default values are given in data lines following the keyword lines. In the data lines we also show names of the corresponding variables, which are referred to in the text.

## 2.1.  General data

This section lists keywords pertaining to definitions of particle numbers, iterations, and phase space.

---

### 2.1..1  Keyword: NUCLIDE

$$86,\ 66 = \texttt{IN\_FIX},\ \texttt{IZ\_FIX}$$

Numbers of neutrons (`IN_FIX`) and protons (`IZ_FIX`) in the nucleus under consideration. Calculations for odd and odd-odd nuclei can be done only either with `IROTAT`=1 or with `IPAIRI`=1 and `IPAHFB`=0. In the latter case, the odd and odd-odd nuclei are described within the BCS blocking approximation with the time-odd potentials neglected.

---

### 2.1.2.  Keyword: ITERATIONS

$$50 = \texttt{NOITER}$$

For `NOITER`>0, the specified number of iterations is performed. Unless `ICONTI`=1 or `IREAWS`=1, the code starts with iteration number 0 by constructing the initial mean fields based on the Nilsson potential. Specifying `NOITER`=0 requests only this initial phase. If the iteration is restarted from the previously recorded potentials (`ICONTI`=1), then the counting of iterations continues from the number of iterations performed in the previous run.

---

### 2.1.3.  Keyword: ITERAT_EPS

$$0.0 = \texttt{EPSITE}$$

---

### 2.1.4.  Keyword: MAXANTIOSC

$$1 = \texttt{NULAST}$$

These two parameters govern the termination of the HF iterations according to the achieved stability of solutions. The stability of the HF energy has been defined in (I-37) as the difference between the total energies calculated from the single-particle energies and from the Skyrme functional. The HF iterations continue until the absolute value of the stability is smaller than `EPSITE` (in MeV) over `NULAST` consecutive iterations. When this condition is fulfilled, iteration procedure terminates and the final results are printed. This allows for an automated adjustment of the number of iterations that are required to achieve a given level of convergence. The number of iterations `NOITER`, can now be set to a large value at which the iterations terminate if a stable solution is not found.

The default value of `EPSITE`=0.0 deactivates this option, and then the code continues up to `NOITER` iterations regardless of the achieved stability. If a non-zero value of `EPSITE` is used, a non-zero value of `NULAST` has to be used too. In practice, a value of `NULAST`=5 prevents an accidental termination of iterations in all cases when the stability energy gradually goes through zero, but the solution is not yet self-consistent.

---

### 2.1.5. Keyword: `PING-PONG`

$$0.0, 3 = \texttt{EPSPNG, NUPING}$$

The code is able to detect the "ping-pong" divergence described in Sec. III-2.6, i.e., the situation when the HF iteration procedure gives oscillating results in every second iteration. Upon continuing the iteration, both sequences of results, i.e., those which correspond to the iteration numbers being even and odd, stay different but perfectly stable, and hence the correct self-consistent solution is never attained.

The code recognizes such a situation by calculating the averages and variances of the stability energy (I-37), separately in the even and in the odd sequences of results, over the last `NUPING` pairs of iterations. The "ping-pong" divergence condition occurs when both variances become a factor `EPSPNG` smaller than the absolute value of the difference of the corresponding averages, i.e., when

$$\Delta(\delta\mathcal{E})_{\text{even}} \quad < \quad \texttt{EPSPNG} \times |\overline{\delta\mathcal{E}}_{\text{even}} - \overline{\delta\mathcal{E}}_{\text{odd}}|, \tag{1a}$$

$$\Delta(\delta\mathcal{E})_{\text{odd}} \quad < \quad \texttt{EPSPNG} \times |\overline{\delta\mathcal{E}}_{\text{even}} - \overline{\delta\mathcal{E}}_{\text{odd}}|, \tag{1b}$$

where

$$\overline{\delta\mathcal{E}}_{\text{even}} \quad = \quad \left( \sum_{p=0}^{\texttt{NUPING}-1} \delta\mathcal{E}_{n-2p} \right) / \texttt{NUPING}, \tag{2a}$$

$$\overline{\delta\mathcal{E}}_{\text{odd}} \quad = \quad \left( \sum_{p=0}^{\texttt{NUPING}-1} \delta\mathcal{E}_{n-2p-1} \right) / \texttt{NUPING}, \tag{2b}$$

and

$$\Delta(\delta\mathcal{E})_{\text{even}} \quad = \quad \left( \sum_{p=0}^{\texttt{NUPING}-1} \left( \delta\mathcal{E}_{n-2p} - \overline{\delta\mathcal{E}}_{\text{even}} \right)^2 \right)^{1/2} / \texttt{NUPING}, \tag{3a}$$

$$\Delta(\delta\mathcal{E})_{\text{odd}} \quad = \quad \left( \sum_{p=0}^{\texttt{NUPING}-1} \left( \delta\mathcal{E}_{n-2p-1} - \overline{\delta\mathcal{E}}_{\text{odd}} \right)^2 \right)^{1/2} / \texttt{NUPING}. \tag{3b}$$

Here, $n$ denotes the number of the last accomplished HF iteration, and $\delta\mathcal{E}_{n'}$ denotes the stability energy (I-37) obtained in the $n'$-th iteration.

The default value of `EPSPNG`=0.0 deactivates this option, and then the code continues up to `NOITER` iterations regardless of the "ping-pong" divergence. If a non-zero value of `EPSPNG` is used, a value of `NUPING`>1 has to be used too. In practice, values of `EPSPNG`=0.01 and `NUPING`=3 allow for an efficient detection of the "ping-pong" divergence condition.

Upon discovering the "ping-pong" divergence, the HF iterations are terminated and a table of absolute values of maximum differences of single-particle observables between the two sequences of iterations is printed, see Sec. III-2.6. These maximum differences are determined for states in each of the charge–parity–signature, charge–simplex, parity, or charge blocks, depending on the conserved symmetries, and separately for particle and hole states. Whenever such a maximum difference is found for a particle state and for a hole state with adjacent indices, such a pair is proposed as a candidate for the diabatic blocking calculation.

---

### 2.1.6. Keyword: <span style="color:blue">CHAOTIC</span>

$$0 = \texttt{NUCHAO}$$

The code is able to detect the "chaotic" divergence which occurs when the HF iterations give results which chaotically vary from one iteration to another one. The code recognizes such a divergence by finding the local maxima $M_k$, $k$=1,2,..., in the sequence of absolute values of the stability energies (I-37), obtained in the entire series of the HF iterations performed. The "chaotic" divergence condition occurs when the code finds `NUCHAO` positive differences $M_k - M_{k-1}$. When this condition occurs, iteration procedure terminates and the final results are printed.

For `NUCHAO`=0 (the default value) the code does not check whether the "chaotic" divergence occurs or not. In practice, a value of `NUCHAO`=5 allows for an efficient detection of the "chaotic" divergence condition. However, for a small value of `NUCHAO` and a small value of `EPSPNG`, the "ping-pong" divergence can sometimes be mistaken for the "chaotic" divergence. If one is interested in the diabatic-blocking data, printed after the "ping-pong" divergence, the recommended value of `NUCHAO`=5 should be increased to 10 or more.

---

### 2.1.7. Keyword: <span style="color:blue">PHASESPACE</span>

$$0,\ 0,\ 0,\ 0 = \texttt{NUMBSP(0,0)},\ \texttt{NUMBSP(1,0)},$$
$$\texttt{NUMBSP(0,1)},\ \texttt{NUMBSP(1,1)}$$

Numbers of the lowest mean-field eigenstates which are kept after the diagonalization of the mean-field Hamiltonians in the four charge–simplex blocks: $(s,q)$=$(+i,n)$, $(-i,n)$, $(+i,p)$, $(-i,p)$. All other eigenstates are discarded. If any of these numbers is equal to zero (the default value), the code sets it equal to the number of neutrons `IN_FIX` (for $q$=n) or protons `IZ_FIX` (for $q$=p). When no symmetry is conserved, the numbers of states kept in charge blocks are determined by sums `NUMBSP(0,0)` + `NUMBSP(1,0)` and `NUMBSP(0,1)` + `NUMBSP(1,1)`

For calculations without pairing, i.e. for `IPAIRI`=0, the user is responsible for using values of `NUMBSP` large enough to accommodate all wave functions which might be useful for the required vacuum and particle-hole configurations, see Sec, II-3.4. In practice, the use of the default values described above is recommended as a safe option. The size of the matrices defined by the `NDSTAT` parameter can be reduced by the user when small values of `NUMBSP` are used.

For calculations with the BCS pairing, i.e. for `IPAIRI`=1 and `IPAHFB`=0, values of `NUMBSP` define the pairing window in which the BCS equations are solved.

For calculations with the HFB pairing, i.e. for `IPAIRI`=1 and `IPAHFB`=1, values of `NUMBSP` define the numbers of canonical states kept after diagonalizing density matrices. They are irrelevant for the pairing window, which is then defined by the cutoff energy `ECUTOF`.

## 2.2.   Interaction

This section lists keywords pertaining to definitions of coupling constants and interaction parameters.

---

### 2.2.1.   Keyword: `SKYRME-SET`

$$\text{SKM* = \texttt{SKYRME}}$$

`CHARACTER*4` acronym of the Skyrme force parameter set. Must start at the 13-th column of the data line. In version (v2.08k), valid acronyms are SIII, SV, SKM*, SKP, SKMP, SKI1, SKO, SKOP, SLY4, SLY5, MSK1, MSK2, MSK3, MSK4, MSK5, MSK6, SKX, SKXC. Other sets of parameters can easily be included in the subroutine `PARAMS`.

---

### 2.2.2.   Keyword: `SKYRME-STD`

$$0, 1, 0, 0, 0 = \texttt{ISTAND,KETA\_J,KETA\_W,KETACM,KETA\_M}$$

Parameters of several standard Skyrme forces are encoded within the program. Calculation for a given standard force can be requested by specifying in the input data file its acronym `SKYRME`. Along with the force parameters, for each force there is encoded information on how the given force should be used, i.e., with which value of the parameter $\hbar^2/2m$, and with which treatment of tensor, spin-orbit, and center-of-mass terms. For `ISTAND`=1, calculations are performed with these features set in the way specific for the given force, and the rest of the switches read on the same line is ignored. For `ISTAND`=0, the switches define nonstandard features in the following way:

- For `KETA_J`=1 the code takes into account in the functional, and for `KETA_J`=0 neglects, the so-called tensor $J^2$ terms. The second option is equivalent to setting in Eq. (I-12) coupling constants $C_t^J$=0 and $C_t^T$=0.

- For `KETA_W`=0 or 1 the code uses the traditional ($W_0' \equiv W_0$) or generalized ($W_0' \neq W_0$) spin-orbit term, respectively, where

$$C_0^{\nabla J} = C_0^{\nabla j} = -\tfrac{1}{2}W_0 - \tfrac{1}{4}W_0', \tag{4a}$$

$$C_1^{\nabla J} = C_1^{\nabla j} = -\tfrac{1}{4}W_0'. \tag{4b}$$

  For forces that use generalized spin-orbit term, option `KETA_W`=0 sets $W_0'$ equal to $W_0$, while for forces that use traditional spin-orbit, option `KETA_W`=1 has no effect. For `KETA_W`=2, strengths $W_0$ and $W_0'$ are set equal to the values of `WO_INP` and `WOPINP`, respectively, which are read under the keyword of `SPIN_ORBIT`. Note that if `KETA_W`=2 is used, and keyword `SPIN_ORBIT` is not specified in the input data file, the default values of `WO_INP` and `WOPINP` supersede those that are encoded within the program for the given Skyrme force.

- For `KETACM`=0 the code uses the traditional one-body center-of-mass correction before variation,

$$E_{\text{c.m.}} \simeq E_{\text{c.m.}}^{\text{dir}} = -\frac{1}{A}\langle \hat{T} \rangle, \tag{5}$$

where $\hat{T} = -\frac{\hbar^2}{2m}\sum_{i=1}^{A} \nabla_i^2$ is the one-body kinetic-energy operator. For `KETACM`=1 the code uses the two-body center-of-mass correction after variation,

$$E_{\text{c.m.}} = -\frac{1}{2mA}\langle \hat{\boldsymbol{P}}^2 \rangle, \tag{6}$$

where $\hat{\boldsymbol{P}} = -i\hbar \sum_{i=1}^{A} \nabla_i$ is the total linear momentum operator. Value of `KETACM`=2 is reserved for a future implementation of the two-body center-of-mass correction before variation. For `KETACM`=3 the center-of-mass correction is neglected.

- For `KETA_M`=0 the code uses the value of $\hbar^2/2m$=20.73620941 MeV fm$^2$, which was encoded in version (v1.75r). For `KETA_M`=1 the code uses the value specific for the given Skyrme force. For `KETA_M`=2 the code uses the value specified in the input data file under keyword `HBAR2OVR2M`.

---

### 2.2.3. Keyword: `HBAR2OVR2M`

$$20.73620941 = \text{HBMINP}$$

Value of the $\hbar^2/2m$ parameter, which is used if `KETA_M`=2 is set under keyword `SKYRME-STD`, and ignored otherwise.

---

### 2.2.4. Keyword: `SPIN_ORBIT`

$$120.0, 120.0 = \text{WO\_INP,WOPINP}$$

Strengths $W_0$ and $W_0'$ of the generalized spin-orbit interaction (4), which are used if `KETA_W`=2 is set under keyword `SKYRME-STD`, and ignored otherwise.

---

### 2.2.5. Keyword: `LANDAU`

$$0, 0.0, 0.0, 0.0, 0.0, 0.0, 0.0 = \text{LANODD,XO\_LAN,X1\_LAN,}$$
$$\text{GO\_LAN,GOPLAN,}$$
$$\text{G1\_LAN,G1PLAN}$$

Three-digit steering switch `LANODD` is used to specify the type of operations performed in order to define selected time-odd coupling constants based on values of the Landau parameters. The switch is followed by values of $x_0$, $x_1$, $g_0$, $g_0'$, $g_1$, and $g_1'$, as given by equations:

$$x_0 = C_0^s[0]/C_0^s[\rho_{\text{sat}}], \tag{7a}$$
$$x_1 = C_1^s[0]/C_1^s[\rho_{\text{sat}}], \tag{7b}$$

$$g_0 = N_0(2C_0^s[\rho_{\text{sat}}] + 2C_0^T \beta \rho_{\text{sat}}^{2/3}), \tag{7c}$$

$$g_0' = N_0(2C_1^s[\rho_{\text{sat}}] + 2C_1^T \beta \rho_{\text{sat}}^{2/3}), \tag{7d}$$

$$g_1 = -2N_0 C_0^T \beta \rho_{\text{sat}}^{2/3}, \tag{7e}$$

$$g_1' = -2N_0 C_1^T \beta \rho_{\text{sat}}^{2/3}, \tag{7f}$$

where $\beta = (3\pi^2/2)^{2/3}$ and

$$N_0 = \pi^{-2} \left(\frac{\hbar^2}{2m}\right)^{-1} \left(\frac{m^*}{m}\right) \left(\frac{3\pi^2 \rho_{\text{sat}}}{2}\right)^{1/3}. \tag{8}$$

For `LANODD`=000, coupling constants are not defined by the Landau parameters, and the rest of parameters read on the same data line is ignored. For `LANODD`=111, Eqs. (7) are solved, and coupling constants $C_0^s[0]$, $C_0^s[\rho_{\text{sat}}]$, $C_1^s[0]$, $C_1^s[\rho_{\text{sat}}]$, $C_0^T$, and $C_1^T$ are determined from the values of $x_0$, $x_1$, $g_0$, $g_0'$, $g_1$, and $g_1'$. For other values of `LANODD`, the following three steps are performed in sequence:

1. *Step one:* If the rightmost digit of `LANODD` is equal to 1 then the coupling constants $C_0^T$ and $C_1^T$ are determined form $g_1$ and $g_1'$, Eqs. (7e) and (7f); otherwise these two coupling constants are determined from the Skyrme force parameters.

2. *Step two:* If the middle digit of `LANODD` is equal to 1 then the coupling constants $C_0^s[\rho_{\text{sat}}]$ and $C_1^s[\rho_{\text{sat}}]$ are determined form $g_0$ and $g_0'$, Eqs. (7c) and (7d); otherwise these two coupling constants are determined from the Skyrme force parameters.

3. *Step three:* If the leftmost digit of `LANODD` is equal to 1 then the coupling constants $C_0^s[0]$ and $C_1^s[0]$ are determined form $x_0$ and $x_1$, Eqs. (7a) and (7b); otherwise these two coupling constants are determined from the Skyrme force parameters.

All combinations of zeros and ones are allowed in `LANODD`.

---

### 2.2.6.  Keyword: `LANDAU-SAT`

$$-1.0, -1.0, -1.0 = \texttt{HBMSAT,RHOSAT,EFFSAT}$$

Values of parameters $\hbar^2/2m$, $m^*/m$, and $\rho_{\text{sat}}$, respectively, which are used when solving Eq. (7). If a negative number is read for any of these parameters, then the program uses the corresponding value calculated from parameters of the given Skyrme force.

---

### 2.2.7.  Keyword: `EVE_SCA_TS`

$$1., 1., 1., 1., 1., 1., 1., 1., 1., 1., 1., 1.$$

$$\texttt{SRHO\_T, SRHO\_S,SRHODT, SRHODS,SLPR\_T, SLPR\_S,}$$
$$\texttt{STAU\_T, STAU\_S,}$$
$$\texttt{SSCU\_T, SSCU\_S,}$$
$$\texttt{SDIV\_T, SDIV\_S}$$

By using this item, the coupling constants corresponding to a given Skyrme parameter set can be arbitrarily scaled. This allows calculations with modified Skyrme functionals. The time-even coupling constants in the total-sum representation,

$$C^\rho_{\text{tot}} = C^\rho_0 - C^\rho_1, \tag{9a}$$

$$C^\rho_{\text{sum}} = 2C^\rho_1, \tag{9b}$$

are multiplied by the 12 numbers from `SRHO_T` to `SDIV_S`. Variables with names ending with `_T` and `_S` multiply the "total" and "sum" coupling constants, respectively. Variables with names containing the acronyms `RHO`, `LPR`, `TAU`, `SCU`, and `DIV`, multiply the coupling constants with superscripts $\rho$, $\Delta\rho$, $\tau$, $J$, and $\nabla J$, respectively, and those with `RHOD` multiply the density-dependent part of $C^\rho$. Similar name convention is used for many other variables in the code `HFODD`.

---

### 2.2.8. Keyword: `ODD_SCA_TS`

```
                1., 1., 1., 1., 1., 1., 1., 1., 1., 1., 1., 1.
                SSPI_T, SSPI_S,SSPIDT, SSPIDS,SLPS_T, SLPS_S,
                                      SCUR_T, SCUR_S,
                                      SKIS_T, SKIS_S,
                                      SROT_T, SROT_S
```

Same as above but for the time-odd coupling constants. Acronyms `SPI`, `LPS`, `CUR`, `KIS`, and `ROT` correspond to coupling constants with superscripts $s$, $\Delta s$, $T$, $j$, and $\nabla j$, respectively, and those with `SPID` correspond to the density-dependent part of $C^s$.

---

### 2.2.9. Keyword: `EVE_SCA_PM`

```
                1., 1., 1., 1., 1., 1., 1., 1., 1., 1., 1., 1.
                SRHO_P, SRHO_M,SRHODP, SRHODM,SLPR_P, SLPR_M,
                                      STAU_P, STAU_M,
                                      SSCU_P, SSCU_M,
                                      SDIV_P, SDIV_M
```

Same as above but for the time-even coupling constants in the isoscalar-isovector representation,

$$C^\rho_0 = \tfrac{1}{2}C^\rho_{\text{sum}} + C^\rho_{\text{tot}}, \tag{10a}$$

$$C^\rho_1 = \tfrac{1}{2}C^\rho_{\text{sum}}. \tag{10b}$$

Variables with names ending with `_P` and `_M` multiply the isoscalar and isovector coupling constants, respectively. The total-sum scaling factors are used first, and the isoscalar-isovector scaling factors are used afterwards.

**2.2.10.   Keyword: `ODD_SCA_PM`**

$$1., 1., 1., 1., 1., 1., 1., 1., 1., 1., 1., 1.$$

`SSPI_P, SSPI_M,SSPIDP, SSPIDM,SLPS_P, SLPS_M,`
`SCUR_P, SCUR_M,`
`SKIS_P, SKIS_M,`
`SROT_P, SROT_M`

Same as above but for the time-odd coupling constants in the isoscalar-isovector representation.

---

**2.2.11.   Keyword: `G_SCALING`**

$$1.0, 1.0 = \texttt{FACTGN, FACTGP}$$

For `IPAIRI`=1 the code HFODD solves the BCS equations with the neutron and proton pairing strengths defined in Ref. [6]. These values can be modified by defining here the multiplicative factors `FACTGN` and `FACTGP` for neutrons and protons, respectively.

---

**2.2.12.   Keyword: `INI_FERMI`**

$$-8.0, -8.0 = \texttt{FERINI(0),FERINI(1)}$$

---

**2.2.13.   Keyword: `INI_DELTA`**

$$1.0, 1.0 = \texttt{DELINI(0),DELINI(1)}$$

For `IPCONT`=0, the HFB calculations that are restarted from previously saved results are performed with new values of the neutron and proton Fermi energies, `FERINI(0)` and `FERINI(1)`, and pairing gaps, `DELINI(0)` and `DELINI(1)`. New values of pairing gaps are implemented by overwriting the old HFB pairing potentials with constant values of `DELINI(0)` and `DELINI(1)` for neutrons and protons, respectively. These constant potentials are ignored after the first iteration, i.e., in the first iteration, mixing of previous and current potentials (see keyword `SLOW-PAIR`) is not performed. A possibility of restarting calculations with nonzero pairing is very useful in case the pairing would have vanished in a former run. For `IPAHFB`=0 or `IPCONT`=0, values of `FERINI` and `DELINI` are ignored.

---

**2.2.14.   Keyword: `FIXDELTA_N`**

$$1.0, 0 = \texttt{DELFIN,IDEFIN}$$

For `IDEFIN`=1, the HFB pairing calculations are performed with a fixed value of the neutron pairing gap equal to `DELFIN`. For `IDEFIN`=0, value of `DELFIN` is ignored and the zero-range density-dependent pairing force (11) is used for neutrons. `IDEFIN`=1 requires `IPAHFB`=1.

---

### 2.2.15. Keyword: FIXDELTA_P

$$1.0, 0 = \text{DELFIP}, \text{IDEFIP}$$

Same as above but for the proton pairing gap.

---

### 2.2.16. Keyword: PAIRNFORCE

$$-200.0, 0.0, 1.0 = \text{PRHO\_N}, \text{PRHODN}, \text{POWERN}$$

Parameters $V_0$, $V_1$, and $\alpha$, respectively, of the zero-range density-dependent pairing force defined by the form-factor:

$$f(\boldsymbol{r}) = V_0 + V_1 \rho^\alpha(\boldsymbol{r}) = V_0 \left( 1 - \left( \frac{\rho(\boldsymbol{r})}{\rho_0} \right)^\alpha \right), \tag{11}$$

which is used in the HFB calculations for neutrons. In case values of $V_1$ and $\alpha$ allow it, the code calculates the value of the saturation density $\rho_0$ that gives the equivalent form of the form-factor (11). In case $\rho_0$ cannot be calculated, the codes set its value to 1; $\rho_0$ is calculated only for the purpose of information, while internally the code uses only the value of $V_1$. For IDEFIN=1 or IPAHFB=0, parameters PRHO_N, PRHODN, POWERN are ignored.

---

### 2.2.17. Keyword: PAIRPFORCE

$$-200.0, 0.0, 1.0 = \text{PRHO\_P}, \text{PRHODP}, \text{POWERP}$$

Same as above but for the proton pairing force.

---

### 2.2.18. Keyword: PAIR_FORCE

$$-200.0, 0.0, 1.0 = \text{PRHO\_T}, \text{PRHODT}, \text{POWERT}$$

Same as above but for the neutron *and* proton pairing force. This keyword is equivalent to using two keywords simultaneously, PAIRNFORCE and PAIRPFORCE, with identical parameters for neutrons and protons.

---

### 2.2.19. Keyword: PAIRINTER

$$-200.0, 0.16, 1.0 = \text{PRHO\_N}, \text{PRHOSN}, \text{POWERN}$$

Parameters $V_0$, $\rho_0$, and $\alpha$, respectively, of the zero-range density-dependent pairing force (11) used in the HFB calculations for neutrons. In case values of $\rho_0$ and $\alpha$ allow it, the code calculates the value of $V_1$ that gives the equivalent form of the form-factor (11). In case $V_1$ cannot be calculated, the code stops. For IDEFIN=1 or IPAHFB=0, parameters PRHO_N, PRHOSN, POWERN are ignored.

---

**2.2.20.    Keyword: `PAIRPINTER`**

$$-200.0,\ 0.16,\ 1.0 = \texttt{PRHO\_P,PRHOSP,POWERP}$$

Same as above but for the proton pairing force.

---

**2.2.21.    Keyword: `PAIR_INTER`**

$$-200.0,\ 0.16,\ 1.0 = \texttt{PRHO\_T,PRHOST,POWERT}$$

Same as above but for the neutron *and* proton pairing force. This keyword is equivalent to using two keywords simultaneously, `PAIRNINTER` and `PAIRPINTER,` with identical parameters for neutrons and protons.

---

**2.2.22.    Keyword: `CUTOFF`**

$$60.0 = \texttt{ECUTOF}$$

The cutoff energy $\bar{e}_{\max}$ for summing up contributions of the HFB quasiparticle states to density matrices, see Sec. IV-2.5. Ignored for `IPAHFB`=0.

## 2.3. Symmetries

This section lists keywords pertaining to selection of conserved and broken symmetries.

---

### 2.3.1. Keyword: SIMPLEXY

$$1 = \text{ISIMPY}$$

For **ISIMPY**=1, calculation with $y$-simplex conserved are performed, while for **ISIMPY**=0 the simplex is broken. Value of **ISIMPY** must be consistent with switches **IPARTY**, **ISIGNY**, **ISIMTY**, and **IROTAT**, see Tables 1 and 2.

---

### 2.3.2. Keyword: SIGNATUREY

$$1 = \text{ISIGNY}$$

For **ISIGNY**=1, calculation with signature conserved are performed, while for **ISIGNY**=0 the signature is broken. Value of **ISIGNY** must be consistent with switches **ISIMPY**, **IPARTY**, **ISIMTX**, and **ISIMTZ**, see Tables 1 and 3.

---

### 2.3.3. Keyword: PARITY

$$-1 = \text{IPARTY}$$

For **IPARTY**=1, calculation with parity conserved are performed, while for **IPARTY**=0 the parity is broken. Value of **IPARTY** must be consistent with switches **ISIMPY** and **ISIGNY**, see Table 1. For **IPARTY**=−1 (the compatibility mode), the code sets **IPARTY**=**ISIGNY** and requires that **ISIMPY**=1

---

### 2.3.4. Keyword: ROTATION

$$1 = \text{IROTAT}$$

For **IROTAT**=1, calculation with time-reversal breaking are performed, while for **IROTAT**=0 the time-reversal symmetry is conserved. In the latter case the calculations are performed only for one value of the simplex, $s=+i$, which gives almost twice shorter CPU times. **IROTAT**=0 is incompatible with providing a non-zero value of the angular frequency or with attempting a calculation for an odd or odd-odd nucleus. **IROTAT**=1 is incompatible with **IPAIRI**=1, unless **IPAHFB**=1, i.e., the rotating solutions can be obtained for the HFB pairing but not for the BCS pairing.

---

### 2.3.5. Keyword: TIMEREVERS

$$0 = \text{ITIREV}$$

Table 1:  Primary set of conserved and nonconserved symmetries allowed in the code HFODD version (v2.08k).

| Option | Symmetries | | | Switches | | |
|--------|-----------|-----------|-----------|--------|--------|--------|
|        | $\hat{S}_y$ | $\hat{R}_y$ | $\hat{P}$ | ISIMPY | ISIGNY | IPARTY |
| P1 | conserved | conserved | conserved | 1 | 1 | 1 |
| P2 | conserved | nonconserved | nonconserved | 1 | 0 | 0 |
| P3 | nonconserved | conserved | nonconserved | 0 | 1 | 0 |
| P4 | nonconserved | nonconserved | conserved | 0 | 0 | 1 |
| P5 | nonconserved | nonconserved | nonconserved | 0 | 0 | 0 |

Table 2:  Secondary set of conserved and nonconserved symmetries allowed in the code HFODD version (v2.08k).

| Option | Symmetries | | | Switches | | |
|--------|-----------|-----------|-----------|--------|--------|--------|
|        | $\hat{S}_y$ | $\hat{T}$ | $\hat{S}_y^T$ | ISIMPY | ITIREV | ISIMTY |
| S1 | conserved | conserved | conserved | 1 | 1 | 1 |
| S2 | conserved | nonconserved | nonconserved | 1 | 0 | 0 |
| S3 | nonconserved | conserved | nonconserved | 0 | 1 | 0 |
| S4 | nonconserved | nonconserved | conserved | 0 | 0 | 1 |
| S5 | nonconserved | nonconserved | nonconserved | 0 | 0 | 0 |

Table 3:  Ternary set of conserved and nonconserved symmetries allowed in the code HFODD version (v2.08k).

| Option | Symmetries | | | Switches | | |
|--------|-----------|-----------|-----------|--------|--------|--------|
|        | $\hat{R}_y$ | $\hat{S}_x^T$ | $\hat{S}_z^T$ | ISIGNY | ISIMTX | ISIMTZ |
| T1 | conserved | conserved | conserved | 1 | 1 | 1 |
| T2 | conserved | nonconserved | nonconserved | 1 | 0 | 0 |
| T3 | nonconserved | conserved | nonconserved | 0 | 1 | 0 |
| T4 | nonconserved | nonconserved | conserved | 0 | 0 | 1 |
| T5 | nonconserved | nonconserved | nonconserved | 0 | 0 | 0 |

For **ITIREV**=1, calculation with time-reversal conserved are performed, while for **ITIREV**=0 this symmetry is broken. This switch is used only as a convenient replacement of switch **IROTAT**; **ITIREV**=1 is equivalent to **IROTAT**=0 and **ITIREV**=0 is equivalent to **IROTAT**=1.

---

### 2.3.6. Keyword: **TSIMPLEX_Y**

$$-1 = \textbf{ISIMTY}$$

For **ISIMTY**=1, calculation with $y$-simplex$^T$ conserved are performed, while for **ISIMTY**=0 this symmetry is broken. Value of **ISIMTY** must be consistent with switches **ISIMPY** and **IROTAT**, see Table 2. For **ISIMTY**=−1 (the compatibility mode), the code sets **ISIMTY** to 1−**IROTAT** and requires that **ISIMPY**=1.

---

### 2.3.7. Keyword: **TSIMPLEXES**

$$1, 1 = \textbf{ISIMTX}, \textbf{ISIMTZ}$$

For **ISIMTX**=1 and/or **ISIMTZ**=1, calculation with conserved symmetries given by $x$-simplex$^T$ and/or $z$-simplex$^T$ are performed, respectively. These symmetries are broken for **ISIMTX**=0 and **ISIMTZ**=0. Values of **ISIMTX** and **ISIMTZ** must be consistent with that of **ISIGNY**, see Table 3.

---

### 2.3.8. Keyword: **TSIMPLEX3D**

$$1, -1, 1 = \textbf{ISIMTX},\textbf{ISIMTY},\textbf{ISIMTZ}$$

This keyword allows to simultaneously input all the three switches that define the three $T$-simplexes. It is equivalent to using keywords **TSIMPLEXES** and **TSIMPLEX_Y** together.

---

### 2.3.9. Keyword: **PAIRING**

$$0 = \textbf{IPAIRI}$$

For **IPAIRI**=0, calculation without pairing correlations (pure HF) are performed, while for **IPAIRI**=1 the pairing correlations are included. In the latter case, for **IPAHFB**=0 or 1 pairing is included within the BCS or HFB method, respectively. **IPAIRI**=1 is incompatible with **IROTAT**=1, unless **IPAHFB**=1, i.e., the rotating solutions can be obtained for the HFB pairing but not for the BCS pairing.

---

### 2.3.10. Keyword: **HFB**

$$0 = \textbf{IPAHFB}$$

For **IPAHFB**=1, calculation with pairing correlations included by solving the HFB equation are performed. **IPAHFB**=1 requires **IPAIRI**=1. **IPAHFB**=1 also requires **ISIMPY**=1, i.e., for broken simplex symmetry, solution of the HFB equation is in version (v2.08k) not yet implemented.

## 2.4.   Configurations with no conserved symmetries

This section lists keywords pertaining to definitions of configurations in the case when no symmetries are conserved, and the single-particle or single-quasiparticle space is divided into two charge blocks (neutrons and protons).

### 2.4.1.   Keyword: PHNONE_NEU

$$1,\ 00,\ 00 = \texttt{NUPAHO,KPNONE,KHNONE}$$

Neutron particle-hole excitations pertaining to the situation when all neutrons are in one common block (no simplex, signature, or parity is conserved). **NUPAHO** is the consecutive number from 1 to 5 (up to five sets of excitations can be specified in separate items). Particles are removed from the **KHNONE**-th state and put in the **KPNONE**-th state. At every stage of constructing excitations the Pauli exclusion principle has to be respected (particle removed from an occupied state and put in an empty state). Values equal zero have no effect. Note that for all neutrons sitting in one common block the reference configuration from which the particle-hole excitations are counted is defined by the total number of neutrons. These particle-hole excitations are ignored unless **ISIMPY**=0, **IPARTY**=0, and **IPAIRI**=0.

### 2.4.2.   Keyword: PHNONE_PRO

$$1,\ 00,\ 00 = \texttt{NUPAHO,KPNONE,KHNONE}$$

Same as above but for the proton particle-hole excitations.

### 2.4.3.   Keyword: DIANON_NEU

$$2,\ 1,\ 0 = \texttt{KPFLIZ,KHFLIZ,KOFLIZ}$$

The diabatic blocking of neutron single-particle configurations in the situation when all neutrons are in one common block. The rules to define the blocking are analogous to those described for the conserved parity, keyword **DIAPAR_NEU**.

### 2.4.4.   Keyword: DIANON_PRO

$$2,\ 1,\ 0 = \texttt{KPFLIZ,KHFLIZ,KOFLIZ}$$

Same as above but for the diabatic blocking of proton single-particle configurations in the situation when all protons are in one common block.

## 2.5. Configurations with conserved simplex

This section lists keywords pertaining to definitions of configurations in the case when simplex is conserved, and the single-particle or single-quasiparticle space is divided into four charge–simplex blocks (neutrons and protons with simplexes $s=\pm i$).

---

### 2.5.1. Keyword: VACSIM_NEU

$$43, 43 = \texttt{KVASIM(0)}, \texttt{KVASIM(1)}$$

Numbers of lowest neutron states occupied in the two simplex blocks, denoted by $(+)$ and $(-)$, of given simplexes, $s=+i$ and $s=-i$, respectively. These numbers define the simplex reference configuration from which the particle-hole excitations are counted. The definitions of simplex reference configuration and excitations are ignored unless ISIMPY=1, ISIGNY=0, and IPAIRI=0.

---

### 2.5.2. Keyword: VACSIM_PRO

$$33, 33 = \texttt{KVASIM(0)}, \texttt{KVASIM(1)}$$

Same as above but for the numbers of proton states.

---

### 2.5.3. Keyword: PHSIMP_NEU

$$1, 00, 00, 00, 00 = \texttt{NUPAHO}, \texttt{KPSIMP}, \texttt{KPSIMM}, \texttt{KHSIMP}, \texttt{KHSIMM}$$

Neutron particle-hole excitations in the simplex blocks. NUPAHO is the consecutive number from 1 to 5 (up to five sets of excitations can be specified in separate items). Particles are removed from the KHSIMP-th state in the $(+)$ block and from the KHSIMM-th state in the $(-)$ block, and put in the KPSIMP-th state in the $(+)$ block and in the KPSIMM-th state in the $(-)$ block. At every stage of constructing excitations the Pauli exclusion principle has to be respected (particle removed from an occupied state and put in an empty state). Values equal zero have no effect. In practice, reasonable excitations can only be constructed by consulting the printed lists of single-particle states with their consecutive numbers in given blocks. These particle-hole excitations are ignored unless ISIMPY=1, ISIGNY=0, and IPAIRI=0.

---

### 2.5.4. Keyword: PHSIMP_PRO

$$1, 00, 00, 00, 00 = \texttt{NUPAHO}, \texttt{KPSIMP}, \texttt{KPSIMM}, \texttt{KHSIMP}, \texttt{KHSIMM}$$

Same as above but for the proton particle-hole excitations.

---

**2.5.5.  Keyword: <span style="color:blue">DIASIM_NEU</span>**

$$2, 2, \quad 1, 1, \quad 0, 0 = \quad \texttt{KPFLIM(0,0)},\texttt{KPFLIM(1,0)},$$
$$\texttt{KHFLIM(0,0)},\texttt{KHFLIM(1,0)},$$
$$\texttt{KOFLIM(0,0)},\texttt{KOFLIM(1,0)}$$

The diabatic blocking of neutron single-particle simplex configurations.  Matrices `KPFLIM` contain the indices of the particle states in the two simplex blocks denoted by $(+)$ and $(-)$, of given simplex values, i.e., $s=+i$ and $-i$, respectively.  Matrices `KHFLIM` contain analogous indices of the hole states, and matrices `KOFLIM` define types of blocking in analogy to those shown in Table 4:

---

**2.5.6.  Keyword: <span style="color:blue">DIASIM_PRO</span>**

$$2, 2, \quad 1, 1, \quad 0, 0 = \quad \texttt{KPFLIM(0,1)},\texttt{KPFLIM(1,1)},$$
$$\texttt{KHFLIM(0,1)},\texttt{KHFLIM(1,1)},$$
$$\texttt{KOFLIM(0,1)},\texttt{KOFLIM(1,1)}$$

Same as above but for the diabatic blocking of proton single-particle simplex configurations.

## 2.6. Configurations with conserved parity

This section lists keywords pertaining to definitions of configurations in the case when parity is conserved, and the single-particle or single-quasiparticle space is divided into four charge–parity blocks (neutrons and protons with parities $\pi = \pm 1$).

---

### 2.6.1. Keyword: VACPAR_NEU

$$44, 42 = \text{KVASIQ(0),KVASIQ(1)}$$

Numbers of lowest neutron states occupied in the two parity blocks, denoted by $(+1)$ and $(-1)$, of given parities, $\pi = +1$ and $\pi = -1$, respectively. These numbers define the parity reference configuration from which the particle-hole excitations are counted. The definitions of parity reference configuration and excitations are ignored unless IPARTY=1, or IPARTY=−1 and ISIMPY=ISIGNY=1. They are also ignored for ISIMPY=1 and IPAIRI=1.

---

### 2.6.2. Keyword: VACPAR_PRO

$$32, 34 = \text{KVASIQ(0),KVASIQ(1)}$$

Same as above but for the numbers of proton states.

---

### 2.6.3. Keyword: PHPARI_NEU

$$1, 00, 00, 00, 00 = \text{NUPAHO,KPSIQP,KPSIQM,KHSIQP,KHSIQM}$$

Neutron particle-hole excitations in the parity blocks. NUPAHO is the consecutive number from 1 to 5 (up to five sets of excitations can be specified in separate items). Particles are removed from the KHSIQP-th state in the $(+1)$ block and from the KHSIQM-th state in the $(-1)$ block, and put in the KPSIQP-th state in the $(+1)$ block and in the KPSIQM-th state in the $(-1)$ block. At every stage of constructing excitations the Pauli exclusion principle has to be respected (particle removed from an occupied state and put in an empty state). Values equal zero have no effect. These particle-hole excitations are ignored unless ISIMPY=0, IPARTY=1, and IPAIRI=0.

---

### 2.6.4. Keyword: PHPARI_PRO

$$1, 00, 00, 00, 00 = \text{NUPAHO,KPSIQP,KPSIQM,KHSIQP,KHSIQM}$$

Same as above but for the proton particle-hole excitations.

---

### 2.6.5.   Keyword: `DIAPAR_NEU`

$$2, 2, \quad 1, 1, \quad 0, 0 = \quad \texttt{KPFLIQ(0,0),KPFLIQ(1,0),}$$
$$\texttt{KHFLIQ(0,0),KHFLIQ(1,0),}$$
$$\texttt{KOFLIQ(0,0),KOFLIQ(1,0)}$$

The diabatic blocking of neutron single-particle parity configurations. Matrices `KPFLIQ` contain the indices of the particle states in the two parity blocks denoted by $(+1)$ and $(−1)$, of given parities, i.e., $\pi=+1$ and $−1$, respectively. Matrices `KHFLIQ` contain analogous indices of the hole states, and matrices `KOFLIQ` define the type of blocking according to definitions shown in Table 4:

Table 4:  Values of parameter `KOFLIQ` that specify type of blocking in case of conserved parity. Analogous values define type of blocking for other conserved symmetries.

| | | |
|---|---|---|
| `KOFLIQ=0`   | $\Longleftrightarrow$ | No diabatic blocking in the given parity block. |
| `KOFLIQ=+1`  | $\Longleftrightarrow$ | The state which has the *larger* $y$-alignment is occupied. |
| `KOFLIQ=−1`  | $\Longleftrightarrow$ | The state which has the *smaller* $y$-alignment is occupied. |
| `KOFLIQ=+2`  | $\Longleftrightarrow$ | The state which has the *larger* $y$-intrinsic spin is occupied. |
| `KOFLIQ=−2`  | $\Longleftrightarrow$ | The state which has the *smaller* $y$-intrinsic spin is occupied. |
| `KOFLIQ=+3`  | $\Longleftrightarrow$ | The state which has the *larger* $x$-alignment is occupied. |
| `KOFLIQ=−3`  | $\Longleftrightarrow$ | The state which has the *smaller* $x$-alignment is occupied. |
| `KOFLIQ=+4`  | $\Longleftrightarrow$ | The state which has the *larger* $x$-intrinsic spin is occupied. |
| `KOFLIQ=−4`  | $\Longleftrightarrow$ | The state which has the *smaller* $x$-intrinsic spin is occupied. |
| `KOFLIQ=+5`  | $\Longleftrightarrow$ | The state which has the *larger* $z$-alignment is occupied. |
| `KOFLIQ=−5`  | $\Longleftrightarrow$ | The state which has the *smaller* $z$-alignment is occupied. |
| `KOFLIQ=+6`  | $\Longleftrightarrow$ | The state which has the *larger* $z$-intrinsic spin is occupied. |
| `KOFLIQ=−6`  | $\Longleftrightarrow$ | The state which has the *smaller* $z$-intrinsic spin is occupied. |
| `KOFLIQ=+7`  | $\Longleftrightarrow$ | The state which has the *larger* $o$-alignment is occupied. |
| `KOFLIQ=−7`  | $\Longleftrightarrow$ | The state which has the *smaller* $o$-alignment is occupied. |
| `KOFLIQ=+8`  | $\Longleftrightarrow$ | The state which has the *larger* $o$-intrinsic spin is occupied. |
| `KOFLIQ=−8`  | $\Longleftrightarrow$ | The state which has the *smaller* $o$-intrinsic spin is occupied. |
| `KOFLIQ=+9`  | $\Longleftrightarrow$ | The state which has the *larger* multipole moment $Q_{20}$ is occupied. |
| `KOFLIQ=−9`  | $\Longleftrightarrow$ | The state which has the *smaller* multipole moment $Q_{20}$ is occupied. |
| `KOFLIQ=+10` | $\Longleftrightarrow$ | The state which has the *larger* multipole moment $Q_{22}$ is occupied. |
| `KOFLIQ=−10` | $\Longleftrightarrow$ | The state which has the *smaller* multipole moment $Q_{22}$ is occupied. |

Here, the $x$-, $y$-, and $z$-alignments or intrinsic spins denote projections of the total angular momentum or spin, respectively, on the $x$, $y$, and $z$ axes. Similarly, the $o$-alignment or $o$-intrinsic spin denotes analogous projections on the direction of the angular frequency $\boldsymbol{\omega}_J$.

Within the diabatic blocking procedure one does not predefine whether the particle or the hole state is occupied (as is the case when the particle-hole excitations are defined, see Section 3.4 of II). In each iteration the code calculates the average alignments (or average intrinsic spins, or average quadrupole moments) of both states (those defined by `KPFLIM` and `KHFLIM`), and occupies that state for which a larger, or a smaller value is obtained. Therefore, the order of both states in the Routhian spectrum is irrelevant.

The user is responsible for choosing the particle-state indices (in `KPFLIM`) only among those corresponding to empty single-particle states, and the hole-state indices (in `KHFLIM`) only among those corresponding to occupied single-particle states.

---

### 2.6.6. Keyword: `DIAPAR_PRO`

$$2, 2, \quad 1, 1, \quad 0, 0 = \texttt{KPFLIQ(0,1)}, \texttt{KPFLIQ(1,1)},$$
$$\texttt{KHFLIQ(0,1)}, \texttt{KHFLIQ(1,1)},$$
$$\texttt{KOFLIQ(0,1)}, \texttt{KOFLIQ(1,1)}$$

Same as above but for the diabatic blocking of proton single-particle parity configurations.

## 2.7.  Configurations with conserved parity and signature

This section lists keywords pertaining to definitions of configurations in the case when parity *and* signature are conserved, and the single-particle or single-quasiparticle space is divided into eight charge–parity–signature blocks (neutrons and protons with parities $\pi=\pm1$ and signatures $r=\pm i$).

---

### 2.7.1.  Keyword: `VACSIG_NEU`

$$22,\ 22,\ 21,\ 21 = \text{\textcolor{red}{KVASIG}}(0,0),\ \text{\textcolor{red}{KVASIG}}(0,1),$$
$$\text{\textcolor{red}{KVASIG}}(1,0),\ \text{\textcolor{red}{KVASIG}}(1,1)$$

Numbers of lowest neutron states occupied in the four parity–signature blocks, denoted by $(+,+)$, $(+,-)$, $(-,+)$, and $(-,-)$, of given (parity,signature) combinations, i.e., $(\pi,r)=(+1,+i)$, $(+1,-i)$, $(-1,+i)$, and $(-1,-i)$, respectively. These numbers define the parity–signature reference configuration from which the particle-hole excitations are counted. The definitions of parity–signature reference configuration and excitations are ignored unless `ISIMPY`=1, `ISIGNY`= 1, and `IPAIRI`=0.

---

### 2.7.2.  Keyword: `VACSIG_PRO`

$$16,\ 16,\ 17,\ 17 = \text{\textcolor{red}{KVASIG}}(0,0),\ \text{\textcolor{red}{KVASIG}}(0,1),$$
$$\text{\textcolor{red}{KVASIG}}(1,0),\ \text{\textcolor{red}{KVASIG}}(1,1)$$

Same as above but for the numbers of proton states.

---

### 2.7.3.  Keyword: `PHSIGN_NEU`

$$1,\ 00,00,00,00,\ 00,00,00,00 = \text{\textcolor{red}{NUPAHO}},$$
$$\text{\textcolor{red}{KPPPSP}},\ \text{\textcolor{red}{KPPPSM}},\ \text{\textcolor{red}{KPPMSP}},\ \text{\textcolor{red}{KPPMSM}},$$
$$\text{\textcolor{red}{KHPPSP}},\ \text{\textcolor{red}{KHPPSM}},\ \text{\textcolor{red}{KHPMSP}},\ \text{\textcolor{red}{KHPMSM}}$$

Neutron particle-hole excitations in the parity–signature blocks. Basic principles are the same as for the excitations in the parity blocks. Particles are removed from the `KHPPSP`-th state in the $(+,+)$ block, from the `KHPPSM`-th state in the $(+,-)$ block, from the `KHPMSP`-th state in the $(-,+)$ block, and from the `KHPMSM`-th state in the $(-,-)$ block, and put in the `KPPPSP`-th state in the $(+,+)$ block, in the `KPPPSM`-th state in the $(+,-)$ block, in the `KPPMSP`-th state in the $(-,+)$ block, and in the `KPPMSM`-th state in the $(-,-)$ block. These particle-hole excitations are ignored unless `ISIMPY`=1, `ISIGNY`=1, and `IPAIRI`=0.

---

### 2.7.4. Keyword: PHSIGN_PRO

```
1, 00,00,00,00, 00,00,00,00 =NUPAHO,
                            KPPPSP, KPPPSM, KPPMSP, KPPMSM,
                            KHPPSP, KHPPSM, KHPMSP, KHPMSM
```

Same as above but for the proton particle-hole excitations.

---

### 2.7.5. Keyword: DIASIG_NEU

```
2, 2, 2, 2,   1, 1, 1, 1,    0, 0, 0, 0 =
KPFLIG(0,0,0),KPFLIG(0,1,0),KPFLIG(1,0,0),KPFLIG(1,1,0),
KHFLIG(0,0,0),KHFLIG(0,1,0),KHFLIG(1,0,0),KHFLIG(1,1,0),
KOFLIG(0,0,0),KOFLIG(0,1,0),KOFLIG(1,0,0),KOFLIG(1,1,0)
```

The diabatic blocking of neutron single-particle parity–signature configurations. Matrices KPFLIG contain the indices of particle states in the four parity–signature blocks denoted by $(+,+)$, $(+,-)$, $(-,+)$, and $(-,-)$, of given (parity,signature) combinations, i.e., $(\pi, r) = (+1,+i)$, $(+1,-i)$, $(-1,+i)$, and $(-1,-i)$, respectively. Matrices KHFLIG contain analogous indices of hole states, and matrices KOFLIG define the type of blocking according to Table 4. Other rules described for the parity case apply here analogously.

---

### 2.7.6. Keyword: DIASIG_PRO

```
2, 2, 2, 2,   1, 1, 1, 1,    0, 0, 0, 0 =
KPFLIG(0,0,1),KPFLIG(0,1,1),KPFLIG(1,0,1),KPFLIG(1,1,1),
KHFLIG(0,0,1),KHFLIG(0,1,1),KHFLIG(1,0,1),KHFLIG(1,1,1),
KOFLIG(0,0,1),KOFLIG(0,1,1),KOFLIG(1,0,1),KOFLIG(1,1,1)
```

Same as above but for the diabatic blocking of proton single-particle parity–signature configurations.

## 2.8. Miscellaneous parameters

This section lists keywords pertaining to definitions of parameters specifying various numerical conditions of the code execution.

---

### 2.8.1. Keyword: MAX_MULTIP

$$2, 4, 4 = \text{NMUCON, NMUCOU, NMUPRI}$$

Maximum multipolarities $\lambda$ of multipole moments used in the code for the constraints, Eq. (15), surface term of the Coulomb field, and printed on the output, respectively. Values not larger than $\lambda=9$ are currently allowed. In case of the conserved parity, only even multipoles are used in the Coulomb field.

---

### 2.8.2. Keyword: MAX_SURFAC

$$0, 0 = \text{NSICON, NSIPRI}$$

Maximum multipolarities $\lambda$ of surface multipole moments, used in the code for the constraints, Eq. (15), and printed on the output, respectively. Values not larger than $\lambda=7$ are currently allowed.

---

### 2.8.3. Keyword: MAX_MAGNET

$$0, 0 = \text{NMACON, NMAPRI}$$

Same as above but for the magnetic moments. In version (v2.08k) constraints on magnetic moments are not yet implemented, which requires NMACON=0. Values not larger than $\lambda=9$ are currently allowed.

---

### 2.8.4. Keyword: COULOMB

$$80, 79, 0.25 = \text{NUMCOU, NUMETA, FURMAX}$$

NUMCOU gives the number of points $N^{\text{Coul}}$, Eq. (I-100), used when summing up the Coulomb Green function. The dimensionless parameter $d$ defining the size of the Coulomb parallelepiped, Eqs. (I-102) and (I-96), is given by NUMCOU*FURMAX. NUMETA gives the order of the Simpson integration of the solid harmonics on the faces of the parallelepiped, see Sec. I-5.

---

### 2.8.5. Keyword: SLOW-DOWN

$$0.5, 0.5 = \text{SLOWEV, SLOWOD}$$

The standard prescription to calculate the HF potential in the next iteration is to mix a given fraction $\epsilon$ of the HF potentials from the previous iteration, with the fraction $1-\epsilon$ of potentials calculated in the current iteration. **SLOWEV** and **SLOWOD** give the values of $\epsilon$ separately for the time-even and time-odd potentials. This mixing slows down the convergence, and is necessary for the convergence of the iterative procedure.

---

### 2.8.6. Keyword: **SLOW-PAIR**

$$0.5 = \text{SLOWPA}$$

Similarly as in the particle-hole channel, **SLOWPA** gives the value of the mixing fraction $\epsilon$ used in the pairing channel, in analogy to **SLOWEV** and **SLOWOD**.

---

### 2.8.7. Keyword: **EPS_HERMIT**

$$1.0\text{E}{-}14 = \text{EPSHER}$$

Numerical precision requested for determining values of Hermite polynomials.

---

### 2.8.8. Keyword: **OPTI_GAUSS**

$$1 = \text{IOPTGS}$$

For **IOPTGS**=1 and **IREAWS**=0, expression $L_\mu = 2N_\mu + 2$ is used to calculate the order of the Gauss-Hermite integration $L_\mu$ from the maximum number $N_\mu$ of HO quanta in the direction $\mu{=}x$, $y$, or $z$, and the input parameters **NXHERM**, **NYHERM**, and **NZHERM** are ignored.

---

### 2.8.9. Keyword: **GAUSHERMIT**

$$18,\ 18,\ 32 = \text{NXHERM},\ \text{NYHERM},\ \text{NZHERM}$$

Orders $L_x$, $L_y$, and $L_z$ of the Gauss-Hermite integration in three Cartesian directions. Must be even. Ignored if **IOPTGS**=0 or **IREAWS**=1.

## 2.9. Parameters of the HO basis

This section lists keywords pertaining to definitions of parameters of the HO basis.

---

### 2.9.1. Keyword: BASIS_SIZE

<div align="center">15, 301, 800.0 = <span style="color:red">NOSCIL, NLIMIT, ENECUT</span></div>

The HO basis is composed of states having not more than $N_0$=NOSCIL quanta in either of the Cartesian directions, and not more than $M$=NLIMIT states in total. The states are added to the basis according to the increasing energy of the deformed harmonic oscillator,

$$\epsilon_{n_x n_y n_z} = \hbar\omega_x(n_x + \tfrac{1}{2}) + \hbar\omega_y(n_y + \tfrac{1}{2}) + \hbar\omega_z(n_z + \tfrac{1}{2}). \tag{12}$$

In case of degenerate HO states (e.g., for an axially deformed HO) the complete multiplets are included, so the actual number of states LDBASE can be slightly larger than NLIMIT. If NLIMIT>0, the cutoff energy ENECUT is ignored; otherwise all states having HO energy smaller then ENECUT are included in the basis, and NLIMIT is ignored.

---

### 2.9.2. Keyword: HOMEGAZERO

<div align="center">1.2 = <span style="color:red">FCHOMO</span></div>

The code uses the standard value of the spherical HO frequency $\hbar\omega_0$=41 MeV/$A^{1/3}$ multiplied by the scaling factor $f$=FCHOMO.

---

### 2.9.3. Keyword: SURFAC_PAR

<div align="center">86, 66, 1.23 = <span style="color:red">INNUMB, IZNUMB, ROPARM</span></div>

The code HFODD calculates parameters of the HO basis, and the zero-iteration Nilsson potential, by defining the standard nuclear surface $\Sigma$,

$$\Sigma: \ R(\theta,\phi) = c(\alpha)\left[1 + \sum_{\lambda=0}^{\lambda_{max}}\sum_{\mu=-\lambda}^{\lambda}\alpha_{\lambda\mu}Y_{\lambda\mu}(\theta,\phi)\right], \tag{13}$$

corresponding to the volume $\frac{4}{3}\pi$*ROPARM$^3$*(INNUMB+IZNUMB), where $c(\alpha)$ is a function of $\alpha_{\lambda\mu}$ such that the volume enclosed by the surface $\Sigma$ does not depend on $\alpha$, i.e., in particular $c(0)$=$R_0$=ROPARM*(INNUMB+IZNUMB)$^{1/3}$.

---

### 2.9.4. Keyword: SURFAC_DEF

<div align="center">2, 0, 0.61 = <span style="color:red">LAMBDA, MIU, ALPHAR</span></div>

The code defines frequencies of the deformed HO in three directions by using relation $\omega_x R_x = \omega_y R_y = \omega_z R_z$, where $R_\mu$ are the lengths of principal axes of the nuclear surface $\Sigma$ (13) defined by real deformation parameters $\alpha_{\lambda\mu}$=`ALPHAR(LAMBDA,MIU)`. The overall factor is defined by $(\omega_x\omega_y\omega_z)^{1/3}$=$\omega_0$.

## 2.10.    Constraints on angular momentum

This section lists keywords pertaining to definitions of constraints on angular momentum. The complete cranking term in the energy density functional is given by:

$$\mathcal{E}^{\text{cran}} = \sum_{a=x,y,z} \left[ -\omega_{J0a}\langle \hat{J}_{0a}\rangle + C_{Ja}\left(\langle \hat{J}_{0a}\rangle - \bar{J}_{0a}\right)^2 - \omega_{J1a}\langle \hat{J}_{1a}\rangle \right] + C_A \left(\frac{\boldsymbol{\omega}_{J0}\times\langle \hat{\boldsymbol{J}}_0\rangle}{\omega_{J0}}\right)^2. \quad (14)$$

### 2.10.1.    Keyword: OMEGAY

$$0.5 = \texttt{OMEGAY}$$

Isoscalar angular frequency $\omega_{J0y}$ in the $y$ direction, see Eq. (14). A non-zero value of OMEGAY requires IROTAT=1 and cannot be used for the BCS pairing, i.e., for IPAIRI=1 and IPAHFB=0.

### 2.10.2.    Keyword: OMISOY

$$0.00 = \texttt{OMISOY}$$

Same as above for the isovector angular frequency $\omega_{J1y}$ in the $y$ direction.

### 2.10.3.    Keyword: OMEGA_XYZ

$$0.00, 0.00, 0.00, 0 = \texttt{OMEGAX, OMEGAY, OMEGAZ, ITILAX}$$

For ITILAX=1, values of the three Cartesian components of the isoscalar angular frequency vector $\boldsymbol{\omega}_{J0}$, see Eq. (14). For ITILAX=0 these values are ignored. ITILAX=1 requires IROTAT=1, ISIMPY=0, and IPAIRI=0. A non-zero value of OMEGAX, OMEGAY, or OMEGAZ requires broken symmetry ISIMTX=0, ISIMTY=0, or ISIMTZ=0, respectively.

### 2.10.4.    Keyword: OMEGA_RTP

$$0.00, 0.00, 0.00, 0 = \texttt{OMERAD, OMETHE, OMEPHI, ITILAX}$$

Same as above but for values of the standard spherical components of the isoscalar angular frequency, i.e., $\omega_{J0r}$, $\omega_{J0\theta}$, and $\omega_{J0\phi}$.

### 2.10.5.    Keyword: OMISO_XYZ

$$0.00, 0.00, 0.00, 0 = \texttt{OMISOX, OMISOY, OMISOZ, ITISAX}$$

For ITISAX=1, values of the three Cartesian components of the isovector angular frequency vector $\boldsymbol{\omega}_{J1}$, see Eq. (14). For ITISAX=0 these values are ignored. ITISAX=1 requires IROTAT=1,

**ISIMPY**=0, and **IPAIRI**=0. A non-zero value of **OMISOX**, **OMISOY**, or **OMISOZ** requires broken symmetry **ISIMTX**=0, **ISIMTY**=0, or **ISIMTZ**=0, respectively.

---

### 2.10.6. Keyword: OMEGA_TURN

$$0 = \texttt{IMOVAX}$$

For **IMOVAX**=1 or $-1$, the isoscalar angular frequency vector $\boldsymbol{\omega}_{J0}$ is in each iteration set in the direction of the total angular momentum $\boldsymbol{J}_0$, or opposite to this direction, respectively, while its length is kept equal to that fixed by the sum of squares of **OMEGAX**, **OMEGAY**, and **OMEGAZ**. For **IMOVAX**=0, vector $\boldsymbol{\omega}_{J0}$ is not changed during the iteration. **IMOVAX**=1 or $-1$ requires **ITILAX**=1, **ITISAX**=0, and **IFLAGA**=0.

---

### 2.10.7. Keyword: SPINCONSTR

$$0.0, 0.0, 0 = \texttt{STIFFI(2)}, \texttt{ASKEDI(2)}, \texttt{IFLAGI(2)}$$

For **IFLAGI**(2)=1, the quadratic constraint on spin is used together with the linear constraint. Values of **STIFFI**(2) and **ASKEDI**(2) correspond respectively to $C_{Jy}$ and $\bar{J}_{0y}$ in Eq. (14). For **IFLAGI**(2)=0, there is no quadratic constraint on spin. This keyword is maintained only for compatibility reasons; in version (v2.08k) of the code HFODD it has been replaced by **SPICON_XYZ**.

---

### 2.10.8. Keyword: SPICON_XYZ

$$0.0, 0.0, 0 = \texttt{STIFFI(1)},\texttt{ASKEDI(1)},\texttt{IFLAGI(1)}$$
$$0.0, 0.0, 0 = \texttt{STIFFI(2)},\texttt{ASKEDI(2)},\texttt{IFLAGI(2)}$$
$$0.0, 0.0, 0 = \texttt{STIFFI(3)},\texttt{ASKEDI(3)},\texttt{IFLAGI(3)}$$

Same as above but for quadratic constraints in the $x$, $y$, and $z$ directions corresponding to **IFLAGI**(1)=1, **IFLAGI**(2)=1, and **IFLAGI**(1)=1, respectively. **IFLAGI**(1)=1, **IFLAGI**(2)=1, or **IFLAGI**(3)=1 requires broken symmetry **ISIMTX**=0, **ISIMTY**=0, or **ISIMTZ**=0, respectively.

---

### 2.10.9. Keyword: SPICON_OME

$$0.0, 0.0, 0 = \texttt{STIFFA}, \texttt{ASKEDA}, \texttt{IFLAGA}$$

For **IFLAGA**=1, the quadratic constraint on the angle between the angular frequency and angular momentum vectors, see Eq. (14). In version (v2.08k) the angle is constrained to zero. Value of **STIFFA** corresponds to $C_A$ in Eq. (14). Value of **ASKEDA** must be set to 0; this variable is reserved for a future implementation of the constraint to a non-zero angle. For **IFLAGA**=0, there is no quadratic constraint on the angle. **IFLAGA**=1 requires **ISIMPY**=0, **IROTAT**=1, **ITILAX**=1, **ITISAX**=0, **IFLAGI**(1)=0, **IFLAGI**(2)=0, **IFLAGI**(3)=0, and a non-zero value of sum of squares of **OMEGAX**, **OMEGAY**, and **OMEGAZ**.

## 2.11. Constraints on multipole moments

This section lists keywords pertaining to definitions of constraints on multipole moments and surface multipole moments. The complete multipole constraint term in the energy density functional is given by:

$$\mathcal{E}^{\mathrm{mult}} = \sum_{\lambda\mu} C_{\lambda\mu} \left( \langle \hat{Q}_{\lambda\mu} \rangle - \bar{Q}_{\lambda\mu} \right)^2 + \sum_{\lambda\mu} C_{\lambda\mu}^S \left( \langle \hat{Q}_{\lambda\mu}^S \rangle - \bar{Q}_{\lambda\mu}^S \right)^2. \tag{15}$$

The multipole and surface multipole moments are defined as

$$Q_{\lambda\mu}(\boldsymbol{r}) = a_{\lambda\mu} r^\lambda Y_{\lambda\mu}^*(\theta, \phi), \tag{16}$$

and

$$Q_{\lambda\mu}^S(\boldsymbol{r}) = a_{\lambda\mu} r^{\lambda+2} Y_{\lambda\mu}^*(\theta, \phi), \tag{17}$$

respectively, where $Y_{\lambda\mu}$ are the standard spherical harmonics in the convention of Ref. [7]. For $\lambda \leq 2$, factors $a_{\lambda\mu}$ are listed in Table 5, while for $\lambda > 2$ $a_{\lambda\mu} = 1$.

Table 5: Adopted definitions of the normalization factors $a_{\lambda\mu}$ and the corresponding multipole moments $Q_{\lambda\mu}$.

| $\lambda$ | $\mu$ | $a_{\lambda\mu}$ | $Q_{\lambda\mu}$ |
|---|---|---|---|
| 0 | 0 | $\sqrt{4\pi}$ | 1 |
| 1 | 0 | $\sqrt{4\pi/3}$ | $z$ |
| 1 | 1 | $-\sqrt{8\pi/3}$ | $x - iy$ |
| 2 | 0 | $\sqrt{16\pi/5}$ | $2z^2 - x^2 - y^2$ |
| 2 | 1 | $-\sqrt{8\pi/15}$ | $zx - izy$ |
| 2 | 2 | $\sqrt{32\pi/5}$ | $\sqrt{3}(x^2 - y^2 - 2ixy)$ |

---

### 2.11.1. Keyword: MULTCONSTR

<div align="center">2, 0, 0.01, 42.0, 1 = LAMBDA, MIU, STIFFQ, QASKED, IFLAGQ</div>

For IFLAGQ=1, the total multipole moment of the given multipolarity $\lambda$ and $\mu$ is constrained. Values of LAMBDA, MIU, STIFFQ, and QASKED correspond respectively to $\lambda$, $\mu$, $C_{\lambda\mu}$, and $\bar{Q}_{\lambda\mu}$ in Eq. (15). For IFLAGQ=0, there is no constraint in the given multipolarity.

---

### 2.11.2. Keyword: SURFCONSTR

<div align="center">2, 0, 0.0, 0.0, 0 = LAMBDA, MIU, STIFFS, SASKED, IFLAGS</div>

For `IFLAGS`=1, the total surface multipole moment of the given multipolarity $\lambda$ and $\mu$ is constrained. Values of `LAMBDA`, `MIU`, `STIFFS`, and `SASKED` correspond respectively to $\lambda$, $\mu$, $C_{\lambda\mu}^S$, and $\bar{Q}_{\lambda\mu}^S$ in Eq. (15). For `IFLAGS`=0, there is no constraint in the given multipolarity.

## 2.12.   Output-file parameters

This section lists keywords pertaining to definitions of parameters specifying character of information printed on the output file.

---

### 2.12.1.   Keyword: PRINT-ITER

$$1, 0, 1 = \text{IPRSTA, IPRMID, IPRSTO}$$

The code prints results for the first, middle, and/or last iteration if the corresponding parameters IPRSTA, IPRMID, and IPRSTO equal 1.

---

### 2.12.2.   Keyword: PRINT-MOME

$$1, 1, 1 = \text{IPRI\_N, IPRI\_P, IPRI\_T}$$

The code prints values of the neutron, proton, and total moments (multipole moments, surface multipole moments, or magnetic moments) only if the corresponding parameters IPRI_N=1, IPRI_P=1, and IPRI_T=1, respectively.

---

### 2.12.3.   Keyword: PRINT-INTR

$$1 = \text{INTRIP}$$

The code prints the values of moments (multipole moments, surface multipole moments, or magnetic moments) and angular momenta in the intrinsic frame only if INTRIP=1.

---

### 2.12.4.   Keyword: EALLMINMAX

$$-12.0, 0.0 = \text{EMINAL, EMAXAL}$$

The code prints tables of single-particle properties for states with values of the single-particle Routhians between EMINAL and EMAXAL. No table is printed unless EMINAL $\leq$ EMAXAL.

---

### 2.12.5.   Keyword: EQUASI_MAX

$$10.0 = \text{EMAXQU}$$

The code prints tables of quasiparticle properties for the HFB states with values of the quasiparticle energies smaller than EMAXQU. No table is printed unless EMAXQU>0.

## 2.13. Files

This section lists keywords that define files used by the code.

---

### 2.13.1. Keyword: REVIEWFILE

$$\text{HFODD.REV} = \text{FILREV}$$

`CHARACTER*68` file name of the review file. Must start at the 13-th column of the data line. The ASCII review file is written after calculating every data set (i.e., once per every **EXECUTE** item), provided **IREVIE**>0. The file is not rewound, so the results for several data sets can be accumulated in a single file. This is so provided the filename **FILREV** is not changed between the **EXECUTE** items. The file contains all relevant parameters and results of calculation in a form suitable for reading by another program. It is meant as an interface to programs which analyze and/or plot the results. The file contains sections defined by keywords (different than keywords used in the input data file described here). The detailed structure of the review file is not documented in the present write up, and can be inferred from inspecting the specimen produced by the sample run, and from the subroutine **REVIEW**.

---

### 2.13.2. Keyword: REVIEW

$$2 = \text{IREVIE}$$

The review file is not written if **IREVIE**=0. For **IREVIE**=2, the table of single-particle properties is included in the review file in addition to other results, which are written for **IREVIE**=1.

---

### 2.13.3. Keyword: REPLAYFILE

$$\text{HFODD.REP} = \text{FILREP}$$

`CHARACTER*68` file name of the replay file. Must start at the 13-th column of the data line. The binary replay file with the name defined in **FILREP** must exist if **ICONTI**=1, and will be read. If the filenames **FILREP** and **FILREC** are identical, the replay file will be subsequently overwritten as a new record file. These feature is implemented to facilitate chaining of jobs which follow one another.

---

### 2.13.4. Keyword: RECORDFILE

$$\text{HFODD.REC} = \text{FILREC}$$

`CHARACTER*68` file name of the record file. Must start at the 13-th column of the data line. If **IWRIRE**=1, binary record file is written after each HF iteration. It contains complete information which allows restarting the iteration in another run of the code. To restart, one has to specify **ICONTI**=1 and provide the name of the file by defining **FILREP**. In case of the computer crash, or upon a successful completion of the given input data set, the record file contains the results of the last performed iteration.  The file is always rewound before it is written, so the results for consecutive iterations do not pile up.

---

### 2.13.5.    Keyword: RECORDSAVE

$$1 = \texttt{IWRIRE}$$

For IWRIRE=1, the record file, is saved on disc after each iteration is completed.  For IWRIRE=0, it is saved only once, after all iterations are completed.  This option is useful on systems where the speed of writing large amounts of data to disc could hamper the system performance.  For IWRIRE=−1, the record file is never saved, and the run cannot be later continued.

---

### 2.13.6.    Keyword: WOODSAFILE

$$\text{WOODS.WFN} = \texttt{FILWOO}$$

CHARACTER*68 file name of the Woods-Saxon file.  Must start at the 13-th column of the data line.  This file is read provided IREAWS=1.  The binary Woods-Saxon file constitutes an interface between the Woods-Saxon code and the code HFODD.  It contains the Woods-Saxon wave functions and numerous other parameters which define the current calculation.  *The parameters read from the Woods-Saxon file overwrite the values provided in the input data file.*  Since the current version of the Woods-Saxon code will be published separately, the feature of starting the iteration from the Woods-Saxon results is not documented in the present write up.

---

### 2.13.7.    Keyword: COULOMFILE

$$\text{HFODD.COU} = \texttt{FILCOU}$$

CHARACTER*68 file name of the Coulomb file containing auxiliary data for a faster calculation of the Coulomb field.  Must start at the 13-th column of the data line.  Parameters NUMCOU, NUMETA, and FURMAX which define the calculations of the Coulomb direct potential (see Secs. I-5 and II-3.5) are usually kept unchanged for the whole series of calculations in one region of nuclei.  Therefore, many Coulomb auxiliary results can be calculated only once, and stored in the file FILCOU.  Handling of this file is determined by the input parameters ICOULI and ICOULO.

---

### 2.13.8.    Keyword: COULOMSAVE

$$0,0 = \texttt{ICOULI, ICOULO}$$

Input parameters ICOULI and ICOULO determine actions pertaining to reading and/or writing of the Coulomb file FILCOU, according to the following table:

| ICOULI | ICOULO | Action |
|:---:|:---:|:---|
| 0 | 0 | neither read nor write the Coulomb file |
| 1 | 0 | read, but do not write the Coulomb file |
| 0 | 1 | do not read, but write the Coulomb file |
| 1 | 1 | use automated handling of the Coulomb file |

The default values of ICOULI=ICOULO=0 ensure that the code HFODD (v2.08k) behaves as that in version (v1.60r). However, unless it is required by special circumstances, a use of the automated handling of the Coulomb file (ICOULI=ICOULO=1) is recommended. Within the automated mode, the code checks whether the Coulomb file exists, and whether it contains data which match the current values of the input parameters NUMCOU, NUMETA, and FURMAX. If this is the case, the code reads the data from the Coulomb file. If this is not the case, the code calculates the Coulomb auxiliary results and stores them in the Coulomb file FILCOU. In the automated mode, the user is informed by appropriate messages printed on the output file about what type of the action has been taken in the given run of the code.

## 2.14.   Starting the iteration

This section lists keywords that pertain to starting and restarting the code.

---

### 2.14.1.   Keyword: **RESTART**

$$0 = \text{ICONTI}$$

For **ICONTI**=1, results stored in the replay file (written as a record file in a previous run) are used to start the iteration. The code assumes (without checking) that the parameters of the HO basis are the same as those used in the previous run. The replay file name should be provided by defining **FILREP**. If the previous run was done with **IREAWS**=1, the current run must also use **IREAWS**=1, and the same Woods-Saxon file must be provided. This so because the Woods-Saxon file contains not only the information about the starting potential (which is ignored for **ICONTI**=1) but also defines the HO basis.

---

### 2.14.2.   Keyword: **CONT_PAIRI**

$$0 = \text{IPCONT}$$

For **IPCONT**=1, results stored in the replay file are used to define the HFB pairing properties in the first iteration; otherwise values read in matrices **FERINI** and **DELINI** are used. When the HFB pairing correlations are taken into account (i.e., for **IPAHFB**=1), and if a smooth restart and continuation of iterations from previously stored results is required, value of **IPCONT**=1 must be used. **IPCONT**=1 is incompatible with either of **IPAHFB**=0 or **ICONTI**=0.

---

### 2.14.3.   Keyword: **CONT_OMEGA**

$$0 = \text{IOCONT}$$

For **IOCONT**=1, the angular momentum vector $\boldsymbol{J}_0$ stored in the replay file is used to define in the first iteration the direction of the angular frequency vector $\boldsymbol{\omega}_{J0}$, see Eq. (14); otherwise values read in **OMEGAX**, **OMEGAY**, and **OMEGAZ** are used (see keyword **OMEGA_XYZ**). **IOCONT**=1 requires **IMOVAX**=1 or $-1$.

---

### 2.14.4.   Keyword: **READ_WOODS**

$$0 = \text{IREAWS}$$

For **IREAWS**=1, the results stored by the previously performed Woods-Saxon calculation are used to start the iteration and to define the HO basis. For **ICONTI**=1 and **IREAWS**=1 the Woods-Saxon file must also be provided, and is used only to define the HO basis.

---

### 2.14.5. Keyword: `NILSSONPAR`

0, –1.175, –0.247, –1.175, –0.352, 11.17, 11.17, 6.28 =
`NILDAT`, `CNILSN`, `DNILSN`, `CNILSP`, `DNILSP`,
`HBANIX`, `HBANIY`, `HBANIZ`

For `IREAWS`=0 and `ICONTI`=0, the code starts the calculation from the Nilsson potential. If `NILDAT`=1, the Nilsson parameters $C$ and $D$ [Ref. [8], Eq. (2.89)] are for neutrons given by `CNILSN` and `DNILSN` and for protons by `CNILSP` and `DNILSP`, while frequencies of the deformed HO are defined by `HBANIX`, `HBANIY`, and `HBANIZ`. If `NILDAT`=0, the Nilsson parameters $C$ and $D$ are defined by Eq. (2.91) and Table 2.3 of Ref. [8], while frequencies of the deformed HO are the same as those of the HO basis defined under keyword `SURFAC_DEF`. In the latter case the Nilsson parameters read from the input file are ignored.

# 3.   OUTPUT FILE

The code writes results on standard FORTRAN output file. The output file begins with the information pertaining to the general parameters of the calculation, then gives information about the starting point of the iteration, provides the convergence report, and finally contains the results calculated at the last iteration. Structure of the output file strongly depends on the type of calculation that are requested in the given run; especially on the assumed conserved and broken symmetries. As a rule, only the information relevant for the specific run is printed. Descriptions and comments printed on the output file are in most cases self-explanatory. In this section we list only those explanations that are not explicitly included in the output file.

- Section **CLASSICAL NUCLEAR SURFACE** lists deformation parameters used to define the nuclear surface from which the basis parameters are derived, see keyword **OMEGA_XYZ**. It also gives the basis parameters such as the **OSCILLATOR FREQUENCIES: HBAROX, HBAROY**, and **HBAROZ** corresponding to $\hbar\omega_x$, $\hbar\omega_y$, and $\hbar\omega_z$.

- Section **PHYSICAL CONSTANTS** gives values of $\hbar c$ in MeV fm (**H_BARC**), of $\hbar c/e^2$ (**HBCOE2**), of the neutron and proton masses in MeV/$c^2$ (**XMASSN** and **XMASSP**), of the kinetic-energy coefficient $\hbar^2/2m$ before (**HBMASS**) and after the center-of-mass correction (**HBMRPA**), of the elementary charge squared $e^2$ in MeV fm (**ECHAR2**), and of the coefficient preceding the integral in the Coulomb exchange energy (**COULEX**). For details see the comments in the **SETBAS** subroutine.

- Section **OSCILLATOR LENGTHS, CONSTANTS**, and **FREQUENCIES** gives the values of $1/b_\mu$, $b_\mu$, and $\hbar\omega_\mu$, respectively, which characterize the HO basis in three Cartesian directions, $\mu=x$, $y$, $z$.

- Section **BASIS CUT-OFF CONTROL PARAMETERS** gives maximum numbers of the HO quanta in three directions **NXMAXX, NYMAXX**, and **NZMAXX** corresponding to $N_x$, $N_y$, and $N_z$, as well as the orders of the Gauss-Hermite quadratures **NXHERM, NYHERM**, and **NZHERM**, corresponding to $L_x$, $L_y$, and $L_z$, see keyword **OPTI_GAUSS**. It also gives the number $M$ of the HO states included in the basis as requested in the input file (**NLIMIT**) and as used in the calculation (**LDBASE**).

- Section **SHAPE OF THE OSCILLATOR-BASIS DIAMOND** gives the numbers of the HO quanta in a given direction for fixed numbers of the HO quanta in both remaining directions. The output is arranged in such a way that the shape of the grid of points $n_x n_y n_z$ is clearly visualized by projections in every of the three directions.

- Section **SKYRME FORCE DEFINITION** lists the name and parameters of the Skyrme force, together with parameters **KETA_J, KETA_W, KETACM**, and **KETA_M** that define the way the given force should be used (see keyword **SKYRME-STD**).

- Section **COEFFICIENTS DEFINING THE SKYRME FUNCTIONAL** gives values of the coupling constants in the Skyrme functional. These values take into account scaling factors, which are printed in the section **SCALING FACTORS**, unless they are all equal to 1.

- Section `CALCULATIONS WITH THE TILTED-AXIS CRANKING` gives values of components of the angular frequency vector and its length. For switches `IMOVAX`=1 or −1 and `IOCONT`=0, these values are used only in the first iteration, and later ignored, because the angular frequency vector is in each iteration readjusted to be aligned or anti-aligned with the angular momentum vector.

- Sections `PARITY/SIGNATURE`, `SIMPLEX`, or `PARITY CONFIGURATIONS` give the vacuum and particle-hole configurations requested in the input file for the corresponding symmetry.

- Section `CONVERGENCE REPORT` gives the list of performed iterations. For each iteration, one line is printed showing the energy, stability energy, average values of the intrinsic quadrupole moment $Q$ and $\gamma$ deformation, total angular momentum, angular frequency, angle between the angular frequency and angular momentum vectors, and total pairing energy.

- Section `SINGLE-PARTICLE PROPERTIES` lists the single-particle states calculated for the Nilsson, Woods-Saxon, or Hartree-Fock Routhian operators. For every state one line is printed which gives the value of the single-particle Routhian, the consecutive numbers in the parity–signature block or simplex block, the quantum numbers $[N, n_z, \Lambda]\Omega$ of the Nilsson asymptotic state which has the largest component in the given state, the average value of the parity operator (in %), the average values of projections of the intrinsic and total angular momenta (in $\hbar$), and their ratio called the $g$-factor. For broken simplex symmetry, three projections of the total angular momentum and intrinsic spin are printed in the first and second line, respectively, for each single-particle state.

- Section `DENSITY INTEGRALS IN THE SKYRME FUNCTIONAL` gives integrals of products of densities, which appear in the Skyrme functional. Terms in the functional are identified by the acronyms described under keywords `EVE_SCA_TS` and `ODD_SCA_TS`.

- Section `CONTRIBUTIONS TO ENERGY IN THE SKYRME FUNCTIONAL` gives the values of various terms which appear in the Skyrme functional. These contributions are the products of the coupling constants and of the density integrals described above. The sums of time-even and time-odd contributions are also printed.

- Section `EULER ANGLES OF THE PRINCIPAL-AXES FRAME` gives the standard Euler angles $\alpha$, $\beta$, and $\gamma$ [7] in degrees, which define the orientation of the intrinsic frame (principal-axes frame) of reference with respect to the original frame.

- Section `MULTIPOLE MOMENTS` gives values of neutron, proton, or mass multipole moments with respect to the original frame of reference. In particular, the $\lambda$=0 moment corresponds to the number of particles. Similarly, section `MULTIPOLE MOMENTS [UNITS: (10 FERMI)^LAMBDA] [INTRINSIC FRAME]` gives analogous values with respect to the intrinsic frame of reference. Whenever the parity symmetry is broken, yet another analogous section gives information on values of multipole moments with respect to the center-of-mass reference frame.

- Sections `ROOT-MEAN-SQUARE AND GEOMETRIC SIZES` gives the rms average values of the radius and of the $x$, $y$, and $z$ coordinates. In order to better visualize the size of the nucleus, the geometric sizes are also calculated by multiplying the rms radius by $\sqrt{5/3}$ and the rms coordinates by $\sqrt{5}$.

- Sections `MAGNETIC MOMENTS [MAGNETON*FERMI^(LAMBDA-1)]` and `MAGNETIC MOMENTS [MAGNETON*FERMI^(LAMBDA-1)] [INTRINSIC FRAME]` give values of magnetic moments with respect to the original and intrinsic frame, respectively.

- Section `ANGULAR MOMENTA` gives for `ISIMPY`=1 the average values of the total and intrinsic neutron, proton, and total angular momentum (in $\hbar$). It also gives the corresponding values and contributions to the first moment of inertia $\mathcal{J}^{(1)}=I/\omega$ (in $\hbar^2$/MeV). For `ISIMPY`=0, this section gives values of neutron, proton, and total projections of the total angular momentum and intrinsic spin on the three Cartesian axes in the original and intrinsic frame, respectively.

- Sections `NEUTRON CONFIGURATIONS` and `PROTON CONFIGURATIONS` give a visual representation of states occupied in the parity–signature blocks, simplex blocks, or parity blocks. The lines denoted by `CONF:` give the configurations requested in the input data set, while those denoted by `VACC:` give the configurations characterizing the given HF state. By comparing the two sets one can verify whether the requested configuration has been obtained, and eventually devise a new configuration to be calculated. The consecutive numbers printed in the horizontal direction correspond to the consecutive numbers in blocks printed in the section `SINGLE-PARTICLE PROPERTIES`. By comparing the two sections one can effectively associate the Nilsson labels with the calculated configurations and also prepare the configuration input data.

- Section `ENERGIES` gives a summary of the energies calculated for the HF state. The kinetic energy, single-particle energy, and pairing energy are printed for neutrons (`NEU`), protons (`PRO`), and all particles (`TOT`). Then, pairing rearrangement energy `P-REARR`, pairing gap `PAIRGAP`, and Fermi energy `E-FERMI` are printed. The Coulomb direct energy (`DIR`) and Coulomb exchange energy (`EXC`) are printed together with their sum (`TOT`). The multipole (`MULT`), surface multipole (`SURF`), and cranking (`SPIN`) constraint energies, are printed together with the corresponding corrections energies (`CORR.`). Then the rearrangement energy is printed followed by the values of the Routhian and the spin-orbit and Skyrme energies, the latter two split in the time-even (`EVE`) and time-odd (`ODD`) contributions. Finally, the total energies are printed as (`SP`) and (`FUN`), respectively, while their difference is printed as the stability (`STAB`).

# 4.   FORTRAN SOURCE FILE

FORTRAN source of code HFODD (v2.08k) is provided in the file `hfodd.f` and can be modified in several places, as described in this section.

## 4.1.   Dimensions of arrays

The code HFODD uses the arrays' dimensions declared through the `PARAMETER` statements. This allows changing the dimensions and adapting the size of the reserved memory to the problem being solved. Whenever too small a dimension is defined, the code aborts with a message indicating the dimension that should be increased. Substantial amount of memory is required only for arrays depending on the following `PARAMETER` values:

`PARAMETER (NDMAIN=21)`
Corresponds to the maximum number of the HO quanta. Should be larger or equal to the input parameter `NOSCIL` defined under keyword `BASIS_SIZE`.

`PARAMETER (NDBASE=507)`
Corresponds to the maximum number of the HO basis states. Should be larger or equal to the input parameter `NLIMIT` defined under keyword `BASIS_SIZE`. It should also be larger or equal to the actual size of the HO basis `LDBASE`, which can be larger than `NLIMIT` in case of degenerate HO states.

`PARAMETER (NDXHRM=33,NDYHRM=33,NDZHRM=45)`
Correspond to the maximum numbers of the HO Gauss-Hermite integration nodes in three Cartesian directions. Should be larger or equal to the input parameters `NXHERM`, `NYHERM`, and `NZHERM`, respectively, that are defined under keyword `GAUSHERMIT` or determined in an optimal way, see keyword `OPTI_GAUSS`.

`PARAMETER (NDSTA=181)`
Corresponds to the maximum numbers of the single-particle states or quasiparticle states, which are kept after diagonalization. Should be larger or equal to the input parameters `NUMBSP(0,0)` + `NUMBSP(1,0)` and `NUMBSP(0,1)` + `NUMBSP(1,1)`, see keyword `PHASESPACE`.

   On vector machines, parameters `NDBASE`, `NDXHRM`, `NDYHRM`, `NDZHRM`, `NDSTAT` and should be odd integers in order to minimize the risk of bank memory conflicts. By the same token, parameter `NDMAIN` should be even, because it defines the matrix dimensions beginning with 0.

## 4.2.   Vectorization properties

As discussed in Sec. I-4.3, the code HFODD has to operate by using seven-fold nested short loops, and this part does not perform well in a vector processor. However, it turns out that the loops can be artificially made longer in such a way that the final CPU time in a vector processor actually becomes much shorter. All the places where this trick has been applied can be identified in the source file by finding the lines beginning with `CVECTOR` and `CSCALAR`, for example:

```
CVECTOR
              DO KZ=0,LAZOXY(NX,NY)+LAZOXY(MX,MY)
```

```
CSCALAR                    DO KZ=NZ+MZ,0,-2
                              RESULT=RESULT+COEF00(KZ,NZ,MZ,3)*T_AUXI(KZ)
                           END DO
```

The line beginning with `CSCALAR` should be made active on a scalar or superscalar machine. The line immediately below the line which begins with `CVECTOR` should be active on a vector machine. The results of calculation do not depend on which version of loops is activated.

## 4.3.  Library subroutines

The code HFODD requires an external subroutine which diagonalizes complex hermitian matrices. Version (v1.60r), see II, has been prepared with an interface to the NAGLIB subroutine F02AXE, and version (v1.75r), see III, with an interface to the LAPACK subroutine ZHPEV. In the present version (v2.08k), both these interfaces remain supported, and can be activated as described in II and III, respectively. However, the recommended interface is now to the LAPACK subroutine ZHPEVX, as described in this section.

In version (v2.08k) we have implemented interface to the LAPACK subroutine ZHPEVX, which can be downloaded (with dependencies) from
http://netlib2.cs.utk.edu/cgi-bin/netlibfiles.pl?filename=/lapack/complex16/zhpevx.f
This subroutine finds not all, but only the lowest eigenvectors, and hence performs calculations in less CPU time; the gain is particularly significant for a large HO basis. Numbers of eigenvectors to be found are defined by the size of the phasespace, see keyword PHASESPACE.

Subroutine ZHPEVX makes use of the BLAS (Basic Linear Algebra Subprograms) library, which can be included in the above downloaded file (unoptimized version). A version optimized for a given computer can be used as well, if it is available in the system.

Subroutine ZHPEVX and its dependencies are in the `REAL*8`/`COMPLEX*16` version, and should be compiled without promoting real numbers to the double precision. On the other hand, the code HFODD itself does require compilation with an option promoting to double precision. Therefore, the code and the ZHPEVX package should be compiled separately, and then should be linked together.

In order to activate the interface to the LAPACK ZHPEVX subroutine, the following modifications of the code HFODD (v2.08k) have to be made:

1. Change everywhere the value of parameter `I_CRAY`=1 into `I_CRAY`=0.

2. Change everywhere the value of parameter `IZHPEV`=0 into `IZHPEV`=2.

3. If your compiler does not support undefined externals, or subroutines called with different parameters, remove calls to subroutines CGEMM, F02AXE, and ZHPEV.

A set of c-shell and ex-editor scripts is provided within the HFODD distribution file, which allows an easy installation, compilation, and execution of the code on a Linux computer running the INTEL© FORTRAN COMPILER.

## 4.4. FORTRAN-90 version

The code HFODD version (v2.08k) is written in FORTRAN-77. However, on several platforms several tested FORTRAN-77 compilers did not perform well with respect to the memory management. In particular, they often distributed large matrices on computer's stack and heap in a wasteful way, and then the available memory could be insufficient for the code to execute, especially for a large HO basis. Therefore, crucial parts of the code have been promoted to FORTRAN-90, with a use of an explicit memory allocation and deallocation. In the FORTRAN source code provided in the file `hfodd.f`, all these FORTRAN-90 features are commented out and inactive. However, very simple modifications of the source code can easily be performed to transform code HFODD to FORTRAN-90. To this end, the user should run an automated editor script which performs the following two operations:

1. Replace all strings "C F90" by spaces, and

2. Remove two lines of code (or three lines of the file) after lines marked by the string: "C IN F90 VERSION PLEASE REMOVE THE TWO FOLLOWING LINES".

The FORTRAN-90 version should be compiled with the "fixed form" non-standard option of the FORTRAN-90 compiler.

A set of c-shell and ex-editor scripts is provided within the HFODD distribution file, and allows performing the above modifications of the FORTRAN source code.

## 4.5. Printing the execution times

After successful completion of execution, the code HFODD prints the table of CPU times spent in principal subroutines, as well as the total execution time. Since implementation of the time-checking functions widely varies between compilers and platforms, in the file `hfodd.f` all references to these functions are commented out and inactive. However, very simple modifications of the source code can easily be performed to activate one of the following options:

1. Replace all strings "CCPUT" by spaces in order to activate a call to the FORTRAN-90 subroutine `CPU_TIME`, which returns CPU execution times.

2. Replace all strings "CETIM" by spaces in order to activate a call to function `ETIME`, which returns CPU execution times.

3. Replace all strings "CUNIX" by spaces in order to activate a call to the UNIX function `TIME`, which returns wall clock execution times.

4. Replace all strings "CSAL" by spaces in order to activate a call to the SALFORD$^{©}$ FORTRAN subroutine `DCLOCK@`, which returns wall clock execution times.

5. Replace all strings "CF32" by spaces in order to activate a call to the MICROSOFT$^{©}$ F32 subroutine `GETTIM`, which returns wall clock execution times.

Other time-checking functions, available to the user, can be analogously called in the subroutine `CPUTIM`.

# 5. TEST RUNS

Several examples of input and output files are provided with the released versions, (v1.60r), (v1.75r), and (v2.08i), of the program HFODD, or added in this user's guide, as listed in Table 6. The input data files `dy152-a.dat` through `dy152-e.dat` present the valid input items of each version and the default or recommended values of the input parameters. The input data files `tb151-a.dat` through `tb151-d.dat` are distributed with version (v1.75r), and provide examples of calculations with diabatic blocking. The input data files `la132-a.dat` and `la132-b.dat` are specific for version (v2.08i), and illustrate various symmetry options of the code. The files `pb128-a.dat` and `sn120-a.dat` are added in this user's guide and give examples of the spherical HF and HFB calculations, respectively. The test runs are described in detail in Section 5.1, and the sample input data files are reprinted in Section 5.2.

Table 6: Input data files supplied with versions (v1.60r), (v1.75r), and (v2.08i) of the program HFODD or added in this user's guide. Each output file listed has been obtained by running version (2.08k) of the code for corresponding input data file.

| Distribution | Input | Contents | Output |
|---|---|---|---|
| (v1.60r) | `dy152-a.dat` | Only keyword `EXECUTE` | `dy152-a.out` |
|  |  | Example of superdeformed HF calculations |  |
|  | `dy152-b.dat` | All input items of (v1.60r), | `dy152-b.out` |
|  |  | default values of input parameters |  |
| (v1.75r) | `dy152-c.dat` | All input items of (v1.75r), | `dy152-c.out` |
|  |  | default values of input items |  |
|  | `dy152-d.dat` | Input items added in (v1.75r), | `dy152-d.out` |
|  |  | recommended values of input items |  |
|  | `tb151-a.dat` | Example of adiabatic calculations | `tb151-a.out` |
|  | `tb151-b.dat` | Example of adiabatic calculations | `tb151-b.out` |
|  | `tb151-c.dat` | Example of diabatic blocking | `tb151-c.out` |
|  | `tb151-d.dat` | Example of diabatic blocking | `tb151-d.out` |
| (v2.08i) | `dy152-e.dat` | All input items of (v2.08i), | `dy152-e.out` |
|  |  | default values of input items |  |
|  | `la132-a.dat` | Example of tilted-axis-cranking calculations | `la132-a.out` |
|  | `la132-b.dat` | Example of tilted-axis-cranking calculations | `la132-b.out` |
| User's | `pb128-a.dat` | Example of spherical HF calculations | `pb208-a.out` |
| guide | `sn120-a.dat` | Example of spherical HFB calculations | `sn120-a.out` |

## 5.1. Descriptions of the test runs

### 5.1.1. Files `dy152-a.dat` through `dy152-e.dat`

As explained in Section 2., all the input data parameters of the code have their default values, which are assumed if the relevant input items are absent in the input data file. Each parameter introduced in a given version is also valid in all later versions, and its default value remains unchanged. Later versions introduce new parameters, but their default values are such that the new features are not active. Thus, the results of running a later version are identical to those of running an earlier version, provided that the values of those new parameters are left at the default. In particular, running each version with its default data set corresponds to identical conditions and gives the same solution. All the input items valid for a given version are also recognized and read by the subsequent ones. In general, running a later version with an input data file prepared for an earlier version leads to the same solution. However, the structure of the output file may be modified from version to version; in particular because later versions may calculate more observables.

Running the code with the default values results in performing 50 iterations for the superdeformed state in $^{152}$Dy at the angular frequency of $\hbar\omega = 0.5\,\mathrm{MeV}$. The pairing correlations are not taken into account. The HO basis is defined by an axially symmetric (with respect to the $z$ axis) nuclear surface with physical deformation typical for superdeformed bands in the $A \sim 150$ region. The parity, $y$-signature, $x$-simplex$^T$ (and their products) are imposed as conserved symmetries, while the time-reversal symmetry is broken in order to describe rotation (about the $y$ axis). The run starts from the Nilsson potential, and an appropriate weak constraint on the quadrupole moment $\langle \hat{Q}_{20} \rangle$ is used to ensure that the iterations fall into the superdeformed minimum.

The input data files `dy152-a.dat` through `dy152-e.dat` define either the default or recommended values of the input parameters for the released versions of the code. They contain either all or some of the valid input items, as described in Table 6. File `dy152-a.dat` contains only keyword `EXECUTE`, and when used as input for any version of the code it results in performing calculations with the default data set. Files `dy152-b.dat`, `dy152-c.dat`, and `dy152e-dat` contain all the valid input items of versions (v1.60r), (v1.75r), and (v2.08i), respectively, and the input data are identical to the default values. File `dy152-d.dat` contains only the input items introduced in version (v1.75r), and the corresponding new parameters are set to recommended values, which differ from the default ones in the following way. The default values are such that the code works identically to version (v1.60r). The parameters `EPSITE`, `NULAST`, `EPSPNG`, and `NUCHAO`, control the termination of the iterations. By default, they are set so that always the number of iterations specified by `NOITER` is performed. The recommended values define an optimal level, at which convergence, "chaotic" divergence, or "ping-pong" divergence are detected to stop the iterations earlier. The differences between the default and recommended values of the variables governing the diabatic blocking are irrelevant because the diabatic blocking is disabled by default. Also by default, the Coulomb file is neither read from nor written to disc. With the recommended values of the parameters `ICOULI` and `ICOULO`, its reading and writing is handled automatically, which saves the CPU time.

Files `dy152-a.dat` through `dy152-e.dat` serve only as suitable patterns to modify the input parameters. However, it should be a good practice to include in the input data file only those input items which modify the input parameters with respect to the default values. In this way, the danger of accidentally modifying some intricate numerical parameters of the code is minimized.

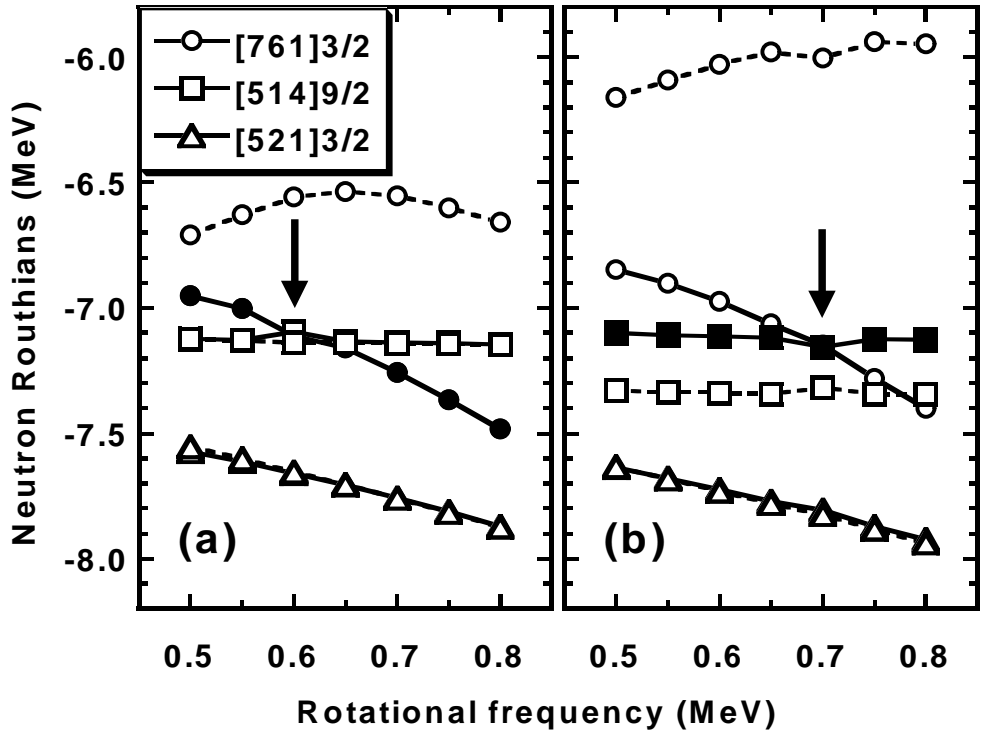

Figure 1: Negative-parity neutron single-particle Routhians in $^{151}$Tb calculated for the [761]3/2($r$=+$i$) (a) and [514]9/2($r$=+$i$) (b) neutron diabatic configuration. Solid and dashed curves denote the $r$=+$i$ and $r$=−$i$ signatures, respectively. The arrows denote the angular frequencies where the converged solutions near the crossing points could not be found.

### 5.1.2.   Files `tb151-a.dat` through `tb151-d.dat`

Files `tb151-a.dat` through `tb151-d.dat` provide data for calculations in which single-particle configurations corresponding to two crossing bands in $^{151}$Tb [9] are followed either adiabatically or diabatically. The pairing correlations are not taken into account. The parity, $y$-signature, $x$-simplex$^T$ (and their products) are imposed as conserved symmetries, while the time-reversal symmetry is broken in order to describe rotation (about the $y$ axis). The angular frequency range from $\hbar\omega$ = 0.50 to 0.80 MeV is scanned with the step of 0.05 MeV. Calculations at $\hbar\omega$ = 0.50 are started from the Nilsson potential, and iterations for each subsequent frequency are restarted from the solution obtained at the preceding frequency. The two considered bands are based on excited physical configurations, in which the last neutron occupies either the [514]9/2($r$=+$i$) or [761]3/2($r$=+$i$) Nilsson level. At $\hbar\omega$ = 0.50 MeV the level [514]9/2($r$=+$i$) is lower in energy, and has number 23−+ (23rd state of $\pi$=−1 and $r$=+$i$), while the level [761]3/2($r$=+$i$) has number 24−+. These two levels cross in the concerned frequency interval so that at $\hbar\omega$ = 0.80 MeV the situation is the opposite. This is illustrated in Figure 1, which is explained in detail below.

In the files `tb151-a.dat` and `tb151-b.dat` it is required that always the level $23-+$ or $24-+$ be occupied, respectively, which amounts to following the configuration adiabatically. If, say, the level $23-+$ is occupied, the solutions at low and high frequencies correspond to physical configurations $[514]9/2(r=+i)$ and $[761]3/2(r=+i)$, respectively, while in the vicinity of the crossing no solution can be found, because the so-called "ping-pong" divergence occurs. This means that the physical states $[514]9/2(r=+i)$ and $[761]3/2(r=+i)$ change their energy order every second iteration, so that each of them once becomes the level $23-+$. Thus, putting the neutron always on the level $23-+$ results in occupying alternately the very different physical states $[514]9/2(r=+i)$ and $[761]3/2(r=+i)$, which cannot lead to convergence.

The code detects the "ping-pong" divergence, stops the iterations, and prints a table that helps identify the crossing orbitals. Such a table is reproduced in Table 7 for the $23-+$ adiabatic configuration at $\hbar\omega = 0.65\,\text{MeV}$. For both protons and neutrons, in each parity–signature block, and separately for the particle and hole states, the code finds the level for which the difference in the angular momentum or intrinsic spin alignment (on the $y$ axis) between the last and last but one iteration is the largest. The indices of those levels in the parity–signature blocks, and the absolute values of the differences are printed in the upper part of the table, independently for the angular momentum and intrinsic spin. If in some symmetry block, the indices of the particle and hole states found in this way differ by one, then it is probable that these two levels cross and cause the "ping-pong" divergence. When such pairs exist, they are listed in the lower part of the table. In the given example, the pair of states $23-+$ and $24-+$ is properly recognized on the basis of both the angular momentum and intrinsic spin. The crossing orbitals, $[514]9/2(r=+i)$ and $[761]3/2(r=+i)$, have markedly different alignments of about $-0.46\hbar$ and $+2.23$, respectively, which corresponds to the values $2.820$ and $2.807\hbar$ given in the table.

If one wishes to follow, say, the configuration $[761]3/2(r=+i)$ diabatically, one should request that, of the levels $23-+$ and $24-+$, the one with physical properties of the state $[761]3/2(r=+i)$ be occupied. One can chose, for instance, the state with higher angular momentum alignment, independent of whether it is lower or higher in energy. In order to diabatically follow the configuration $[514]9/2(r=+i)$, one can chose the state with lower alignment. The corresponding runs are defined in the files `tb151-c.dat` and `tb-151-d.dat`. Indeed, in each case solutions corresponding to the same physical configuration is found at each angular frequency. Moreover, the "ping-pong" divergence does not occur, even very close to the crossing frequency. Figure 1 shows the negative-parity neutron single-particle Routhians obtained from the calculations of files `tb151-c.dat` and `tb-151-d.dat`.

### 5.1.3. Files `la132-a.dat` and `la132-b.dat`

File `la132-a.dat` defines two consecutive runs, which lead to a planar solution [10] in the triaxially deformed nucleus $^{132}$La for $\hbar\omega = 0.25\,\text{MeV}$. The pairing correlations are not taken into account. Since the calculations involve the tilted-axis cranking, a spherically invariant HO basis (with the three HO frequencies equal and comprising only entire HO shells) is used in order not to favor any orientation of the solution in space. The first run is performed at $\hbar\omega = 0$ with the parity, $y$-signature, $x$-simplex$^T$ (and their products) conserved, and with the time-reversal broken in order to describe odd numbers of neutrons and protons. Due to the presence of the odd nucleons, even for $\hbar\omega=0$ one obtains a non-zero angular momentum, which is oriented along the $y$ axis. The run starts from the Nilsson potential, and in 94 iterations finds

Table 7: Example of the output printed when the "ping-pong" divergence is found, see text.

```
*******************************************************************************
*                                                                             *
*   TABLE BELOW GIVES THE MAXIMUM ABSOLUTE VALUES OF DIFFERENCES OF ALIGNMENT  *
*   BETWEEN THE LAST TWO ITERATIONS. IT MAY SERVE AS A GUIDE TO SELECT STATES  *
*   WHICH EXCHANGE WAVE FUNCTIONS IN EVERY SECOND ITERATION.  SUCH STATES MAY  *
*   BE CANDIDATES FOR A CALCULATION WITH DIABATICALLY BLOCKED  CONFIGURATIONS  *
*                                                                             *
*******************************************************************************
*                     *                                                       *
*    BLOCKS           * ANGULAR MOMENTUM ALIGNMENT    INTRINSIC SPIN ALIGNMENT *
*    ------           * --------------------------    ------------------------ *
*                     *   PARTICLES      HOLES          PARTICLES      HOLES    *
*  ISO. PAR. SIG.     * INDEX VALUE   INDEX VALUE      INDEX VALUE   INDEX VALUE *
*                     *                                                       *
*******************************************************************************
*                     *                                                       *
* NEUT   +    +       *   38  0.221    13  0.018        24  0.023    13  0.016  *
* NEUT   +    -       *   28  0.163     7  0.041        33  0.014     7  0.032  *
* NEUT   -    -       *   39  0.095    13  0.198        44  0.015    14  0.150  *
* NEUT   -    +       *   24  2.820    23  2.807        24  0.164    23  0.167  *
* PROT   +    +       *   24  0.255    13  0.149        25  0.129    14  0.122  *
* PROT   +    -       *   27  1.733    13  0.062        21  0.319     8  0.020  *
* PROT   -    -       *   32  0.119    10  0.530        32  0.029    10  0.065  *
* PROT   -    +       *   24  2.649    15  0.082        24  0.153    10  0.030  *
*                     *                                                       *
*******************************************************************************
*                                                                             *
*   FOLLOWS THE LIST OF CANDIDATE CONFIGURATIONS FOR THE DIABATIC BLOCKING     *
*                                                                             *
*******************************************************************************
* NEUT   -    +       *   24  2.820    23  2.807                               *
* NEUT   -    +       *                                 24  0.164    23  0.167  *
*******************************************************************************
```

a triaxial solution, whose shape is forced by using appropriate constraints for the quadrupole moments $\langle \hat{Q}_{20} \rangle$ and $\langle \hat{Q}_{22} \rangle$. The constraints are set so that this solution nearly corresponds to the unconstrained minimum, and are used only to prevent the iterations from falling into other local minima. These constraints are chosen so that the short, medium, and long axes of the mass distribution coincide with the $y$, $x$, and $z$ axes, respectively. For the concerned solution, the single-particle configuration defined in the input data file in terms of numbers of particles in the parity–signature blocks physically corresponds to the odd proton particle occupying the lowest substate of the $h_{11/2}$ orbital and the odd neutron hole left on the highest $h_{11/2}$ substate.

The second run starts from the solution obtained in the first run, and performs calculations with all symmetries broken except parity, with multipole constraints released, and for the angular frequency of $\hbar\omega = 0.25\,\text{MeV}$. The single-particle configuration specified in terms of parity blocks is obtained from the configuration for the first run simply by adding the numbers of particles in the two signature sub-blocks of each parity block. All the three components of the initial angular frequency vector are non-zero, $\hbar\omega_x = \hbar\omega_y = \hbar\omega_z = (0.25/\sqrt{3})$ MeV, which breaks all symmetries except parity in the first iteration, thus making the calculations otherwise symmetry-unrestricted. If a symmetry was not broken in the first iteration, it would artificially remain conserved as governed by the theorem of self-consistent symmetries [8].

The Kerman-Onishi condition [11] states that the angular momentum and angular frequency vectors must be parallel in self-consistent solutions. Standard iteration, with `IMOVAX`=0, requires several thousand iterations for the nucleus to turn appropriately in space to achieve the alignment of these vectors below the angle of 0.01°. On the other hand, iteration with `IMOVAX`=1, as shown in the example of file `la132-a.dat`, achieves the alignment of 0.000001° within 92 iterations. In the solution resulting from the second run, the angular momentum vector has non-zero components only on the $y$ and $z$ axes of the intrinsic frame, which are the short and long axes of the triaxial mass distribution. The calculations of the file `la132-a.dat` are performed in two steps to get a better control over what is going on and to avoid divergence or falling into undesired minima.

The input data provided in the file `la132-b.dat` are similar to those of the file `la132-a.dat`, except that the second run is performed with the $x$-simplex$^T$ conserved. In this way, the angular momentum and angular frequency vectors are confined within the plane $y$-$z$, which is also a principal plane of the mass distribution. After the first run, this plane contains the short and long axes. The components of the initial angular frequency vector are set to $\hbar\omega_y = \hbar\omega_z = (0.25/\sqrt{2})$ MeV and $\hbar\omega_x$=0. Since the planar solution in question corresponds to non-equal projections of the angular momentum on the short and long axes, in order to satisfy the Kerman-Onishi condition, the nucleus still has to turn in space about the $x$ axis. For `IMOVAX`=0, this process is also very slow, while for `IMOVAX`=1, as in file `la132-b.dat`, it takes again only 92 iterations. The $x$-simplex$^T$-imposing solution obtained by using the file `la132-b.dat` differs from the symmetry-unrestricted result of the file `la132-a.dat` only by orientation in space; apart from that, the values of all observables are identical. This shows that in the considered self-consistent solution the $x$-simplex$^T$ symmetry is spontaneously conserved.

### 5.1.4. File `pb208-a.dat`

The input data file `pb208-a.dat` defines two consecutive runs to perform precise calculations for the spherical ground state of the doubly-magic nucleus $^{208}$Pb in a large basis of 26 HO shells. The pairing correlations are not taken into account. Conservation of all the symmetries considered in the code (the group $D_{2h}^T$) is imposed. For the spherical shape, the specified parity–signature configuration corresponds to occupation of the entire magic shells, up to the proton and neutron magic numbers 82 and 126, respectively. A spherical basis is used, i.e., with all the three characteristic HO frequencies equal ($\hbar\omega_x = \hbar\omega_y = \hbar\omega_z = \hbar\omega_0$) and comprising only the entire HO shells. The so-called physical value of $\hbar\omega_0 = f \times 41$ MeV$/A^{1/3}$ with $f = 1.2$ is taken, following the default settings of the code. To speed up the calculations, the program HFODD uses a diagonalization algorithm that can reduce the computational effort by finding less eigenstates than the matrix size (the lowest ones). In the present example, in each charge–simplex block only one extra state is calculated as compared to the number of occupied states in that block, as governed by the input parameter `NUMBSP`. This is far less than the code's default, and such a way of proceeding is possible only because no particle-hole excitations into higher states are required. Also for the sake of saving time, a solution in a small basis of $N_0$=12 HO shells is obtained first, starting from the Nilsson potential, and then serves as a good starting point for the time-consuming calculations with $N_0$=26. The number of states in the spherical basis comprising the HO shells from 0 to $N_0$ equals $M=(N_0+1)(N_0+2)(N_0+3)/6$, and hence the numbers of the HO basis states (for each projection of the intrinsic spin) equal $M$=455 and $M$=3654, respectively for $N_0$=12 and $N_0$=26. Since the code HFODD can restart iterations only if the number of Gauss-Hermite integration points is preserved, both runs are performed with 54 points in each of the Cartesian directions, which is excessive for $N_0$=12, but optimal for $N_0$=26. For $N_0$=26, the 54 points ensure exact integration of all matrix elements apart from those of the Coulomb interaction and density-dependent term.

The energies of $^{208}$Pb obtained from calculations with $N_0$=12 and $N_0$=26 equal $-1630.638$ and $-1635.531$ MeV, respectively. The latter value is very close to the result of $-1635.666$ MeV obtained by using a very precise one-dimensional spherical code working in spatial coordinates [12]. Although the $N_0$=12 HO basis is evidently too small for a precise determination of the energy, other observables may not be so sensitive to the HO basis size. For instance, the mass root-mean-square radii obtained for $N_0$=12 and $N_0$=26 equal 5.5540 and 5.5554 fm, respectively, the result of coordinate-space calculations being 5.5550 fm.

### 5.1.5. File `sn120-a.dat`

The input data file `sn120-a.dat` provides an example of the HFB calculations for the spherical ground state in $^{120}$Sn. All symmetries considered in the code (the group $D_{2h}^T$) are imposed, except the particle-number symmetry, which is broken in order to describe the pairing correlations. A spherical basis is used. The pairing correlations are included within the full HFB method, and the zero-range density-dependent interaction given by the form-factor (11) is used in the particle-particle channel. The particle and pairing densities are obtained by summing up contributions from the quasiparticle states with the equivalent-spectrum energies not exceeding $\bar{\epsilon}_{max}$=60 MeV, which is the code's default pairing window. In the form-factor (11), the power $\alpha$=1 and the saturation density $\rho_0$=0.32 fm$^{-3}$ are taken. The interaction strength, $V_0$=$-285.88$ MeV fm$^3$, was adjusted for the given cutoff energy, $\bar{\epsilon}_{max}$, to reproduce the value of $\tilde{\Delta}_N$=1.245 MeV for the average neutron pairing gap in $^{120}$Sn. The same values of $\alpha$, $\rho_0$ and $V_0$ are taken for both neutrons and protons. The Fermi energy of $-8$ MeV and the pairing gap of 1 MeV are used in the first iteration to construct the HFB Hamiltonian, while the initial mean field is given by the Nilsson potential.

Since $^{120}$Sn is a proton-magic nucleus, the proton pairing decreases during the iteration and vanishes in the HFB solution. As expected, one obtains a significant thickness of the neutron skin (difference between the neutron and proton root-mean-square radii), which in the present case equals 0.14 fm.

## 5.2. Listings of the test input data files

### 5.2.1. File `dy152-a.dat`

EXECUTE

## 5.2.2.  File dy152-b.dat

```
                        ---------- General data  ----------
NUCLIDE
          86    66
ITERATIONS
          50
                        ---------- Interaction  -----------
SKYRME-SET
          SKM*
EVE_SCA_TS  RHO        RHOD       LPR       TAU       SCU       DIV
          1. 1.      1. 1.      1. 1.     1. 1.     1. 1.     1. 1.
ODD_SCA_TS  SPI        SPID       LPS       CUR       KIS       ROT
          1. 1.      1. 1.      1. 1.     1. 1.     1. 1.     1. 1.
EVE_SCA_PM  RHO        RHOD       LPR       TAU       SCU       DIV
          1. 1.      1. 1.      1. 1.     1. 1.     1. 1.     1. 1.
ODD_SCA_PM  SPI        SPID       LPS       CUR       KIS       ROT
          1. 1.      1. 1.      1. 1.     1. 1.     1. 1.     1. 1.
G_SCALING
          1.0  1.0
                        ---------- Symmetries  ------------
ROTATION
          1
SIGNATUREY
          1
TSIMPLEXES
          1      1
PAIRING
          0
                        ---------- Configurations  --------
VACSIM_NEU      SIMP  SIMM
                43    43
VACSIM_PRO      SIMP  SIMM
                33    33
PHSIMP_NEU         PART              HOLE
          1   00    00          00    00
PHSIMP_PRO         PART              HOLE
          1   00    00          00    00
VACSIG_NEU      PPSP  PPSM  PMSP  PMSM
                22    22    21    21
VACSIG_PRO      PPSP  PPSM  PMSP  PMSM
                16    16    17    17
PHSIGN_NEU         PARTICLES                HOLES
          1   00    00    00    00        00    00    00    00
PHSIGN_PRO         PARTICLES                HOLES
          1   00    00    00    00        00    00    00    00
                        ---------- Numerical data  --------
MAX_MULTIP
```

```
          2      4      4
COULOMB
         80     79   0.25
SLOW-DOWN
         0.5  0.5
EPS_HERMIT
         1.00E-14
OPTI_GAUSS
         1
GAUSHERMIT
         18     18     32
                        ---  Parameters of the HO basis  ---
BASIS_SIZE
         15    301   800.
HOMEGAZERO
         1.2
SURFAC_PAR
         86     66   1.23
SURFAC_DEF
          2      0   0.61
SURFAC_DEF
          4      0   0.10
                       ----------  Constraints  ----------
OMEGAY
         0.50
MULTCONSTR
          2      0   0.01   42.   1
SPINCONSTR
                    0.00    0.   0
                        ----  Output-file  parameters  ----
PRINT-ITER
          1      0      1
EALLMINMAX
         -12.   0.
                       -------------  Files  -------------
REVIEWFILE
         HFODD.REV
REVIEW
         2
RECORDFILE
         HFODD.REC
REPLAYFILE
         HFODD.REP
WOODSAFILE
         WOODS.WFN
                       -----  Starting the iteration  ----
RESTART
         0
```

```
READ_WOODS
          0
NILSSONPAR
          0   -1.175 -0.247   -1.175 -0.352    11.170  11.170  6.280
                      ---------- Calculate  -----------
EXECUTE
                      ---------- Terminate  -----------
ALL_DONE
```

### 5.2.3. File dy152-c.dat

```
=======================================================================
|   This file (dy152-c.dat) contains the input data for the code HFODD.   |
| Values of all parameters are identical to the default values assumed by |
|            the code when no input data are provided.            |
=======================================================================
                   ----------  General data  ----------
NUCLIDE
          86    66
ITERATIONS
          50
ITERAT_EPS
          0.0
MAXANTIOSC
          1
PING-PONG
          0.0   3
CHAOTIC
          0
PHASESPACE
          0      0      0      0
                   ----------  Interaction  -----------
SKYRME-SET
          SKM*
EVE_SCA_TS  RHO        RHOD      LPR       TAU       SCU       DIV
          1.  1.    1.  1.    1.  1.    1.  1.    1.  1.    1.  1.
ODD_SCA_TS  SPI        SPID      LPS       CUR       KIS       ROT
          1.  1.    1.  1.    1.  1.    1.  1.    1.  1.    1.  1.
EVE_SCA_PM  RHO        RHOD      LPR       TAU       SCU       DIV
          1.  1.    1.  1.    1.  1.    1.  1.    1.  1.    1.  1.
ODD_SCA_PM  SPI        SPID      LPS       CUR       KIS       ROT
          1.  1.    1.  1.    1.  1.    1.  1.    1.  1.    1.  1.
G_SCALING
          1.0  1.0
                   ----------  Symmetries  ------------
ROTATION
          1
SIGNATUREY
          1
TSIMPLEXES
          1      1
PAIRING
          0
                   ----------  Configurations  --------
VACSIM_NEU            SIMP SIMM
                      43    43
VACSIM_PRO            SIMP SIMM
```

```
                        33   33
PHSIMP_NEU              PARTICLES                HOLES
           1           00   00                   00   00
PHSIMP_PRO             PARTICLES                 HOLES
           1           00   00                   00   00
VACSIG_NEU        PPSP PPSM PMSP PMSM
                   22   22   21   21
VACSIG_PRO        PPSP PPSM PMSP PMSM
                   16   16   17   17
PHSIGN_NEU             PARTICLES                 HOLES
           1   00   00   00   00       00   00   00   00
PHSIGN_PRO             PARTICLES                 HOLES
           1   00   00   00   00       00   00   00   00
DIASIM_NEU             PARTICLES            HOLES            TYPE
                   02   02                 01   01            0  0
DIASIM_PRO             PARTICLES            HOLES            TYPE
                   02   02                 01   01            0  0
DIASIG_NEU             PARTICLES            HOLES            TYPE
               02   02   02   02       01   01   01   01   0  0  0  0
DIASIG_PRO             PARTICLES            HOLES            TYPE
               02   02   02   02       01   01   01   01   0  0  0  0
                    ----------  Numerical data  --------
MAX_MULTIP
           2     4     4
COULOMB
          80    79   0.25
SLOW-DOWN
          0.5  0.5
EPS_HERMIT
          1.00E-14
OPTI_GAUSS
          1
GAUSHERMIT
          18    18    32
                      ---  Parameters of the HO basis  ---
BASIS_SIZE
          15   301   800.
HOMEGAZERO
          1.2
SURFAC_PAR
          86    66   1.23
SURFAC_DEF
          2     0   0.61
SURFAC_DEF
          4     0   0.10
                    ----------  Constraints  ----------
OMEGAY
          0.50
```

```
MULTCONSTR
            2      0   0.01   42.   1
SPINCONSTR
                    0.00    0.   0
                        ----  Output-file  parameters  -----
PRINT-ITER
            1      0      1
EALLMINMAX
            -12.  0.
                        ------------- Files  --------------
REVIEWFILE
            HFODD.REV
REVIEW
            2
RECORDFILE
            HFODD.REC
REPLAYFILE
            HFODD.REP
WOODSAFILE
            WOODS.WFN
COULOMFILE
            HFODD.COU
COULOMSAVE
            0      0
                        -----  Starting the iteration  -----
RESTART
            0
READ_WOODS
            0
NILSSONPAR
            0   -1.175 -0.247   -1.175 -0.352     11.170  11.170  6.280
                        ----------  Calculate  -------------
EXECUTE
                        ----------  Terminate  -------------
ALL_DONE
```

## 5.2.4.  File dy152-d.dat

```
=======================================================================
|   This file (dy152-d.dat) contains the input data for the code HFODD.  |
|     Only keywords introduced after version (1.60r) are included here.   |
=======================================================================
                    ----------  General data  ----------
ITERAT_EPS
          0.0001
MAXANTIOSC
          5
PING-PONG
          0.01    3
CHAOTIC
          5
PHASESPACE
          0        0         0        0
                    ----------  Configurations  --------
DIASIM_NEU           PARTICLES                HOLES              TYPE
               00    00               00    00              0  0
DIASIM_PRO           PARTICLES                HOLES              TYPE
               00    00               00    00              0  0
DIASIG_NEU           PARTICLES                HOLES              TYPE
          00  00  00  00         00  00  00  00      0  0  0  0
DIASIG_PRO           PARTICLES                HOLES              TYPE
          00  00  00  00         00  00  00  00      0  0  0  0
                    -------------  Files  --------------
COULOMFILE
          HFODD.COU
COULOMSAVE
          1       1
                    ----------  Calculate  -------------
EXECUTE
                    ----------  Terminate  -------------
ALL_DONE
```

### 5.2.5. File dy152-e.dat

```
=========================================================================
| This file (dy152-e.dat) contains the input data for the code HFODD.   |
| Values of all parameters are identical to the default values assumed by |
| the code when no input data are provided.                             |
=========================================================================
                    ----------  General data  ----------
NUCLIDE
            86    66
ITERATIONS
            50
ITERAT_EPS
            0.0
MAXANTIOSC
            1
PING-PONG
            0.0   3
CHAOTIC
            0
PHASESPACE
            0       0       0       0
                    ----------  Interaction  -----------
SKYRME-SET
            SKM*
SKYRME-STD
            0       1   0   0   0
HBAR2OVR2M}
            20.73620941
SPIN_ORBIT
            120.0 120.0
LANDAU
            0     0.0  0.0  0.0  0.0  0.0  0.0
LANDAU-SAT
            -1.0   -1.0   -1.0
EVE_SCA_TS  RHO         RHOD        LPR       TAU       SCU       DIV
            1. 1.       1. 1.       1. 1.     1. 1.     1. 1.     1. 1.
ODD_SCA_TS  SPI         SPID        LPS       CUR       KIS       ROT
            1. 1.       1. 1.       1. 1.     1. 1.     1. 1.     1. 1.
EVE_SCA_PM  RHO         RHOD        LPR       TAU       SCU       DIV
            1. 1.       1. 1.       1. 1.     1. 1.     1. 1.     1. 1.
ODD_SCA_PM  SPI         SPID        LPS       CUR       KIS       ROT
            1. 1.       1. 1.       1. 1.     1. 1.     1. 1.     1. 1.
G_SCALING
            1.0  1.0
INI_FERMI
            -8.0   -8.0
INI_DELTA
```

```
          1.0      1.0
FIXDELTA_N
          1.0        0
FIXDELTA_P
          1.0        0
PAIRNFORCE
          -200.0  0.0  1.0
PAIRPFORCE
          -200.0  0.0  1.0
PAIR_FORCE
          -200.0  0.0  1.0
PAIRNINTER
          -200.0 0.16  1.0
PAIRPINTER
          -200.0 0.16  1.0
PAIR_INTER
          -200.0 0.16  1.0
CUTOFF
          60.0
                        ----------  Symmetries  -----------
SIMPLEXY
          1
SIGNATUREY
          1
PARITY
          -1
ROTATION
          1
TIMEREVERS
          0
TSIMPLEX_Y
          -1
TSIMPLEXES
          1       1
TSIMPLEX3D
          1      -1       1
PAIRING
          0
HFB
          0
                        ----------  Configurations  --------
PHNONE_NEU             PARTICLES              HOLES
          1              00                    00
PHNONE_PRO             PARTICLES              HOLES
          1              00                    00
VACSIM_NEU            SIMP SIMM
                       43   43
VACSIM_PRO            SIMP SIMM
```

```
                         33   33
PHSIMP_NEU              PARTICLES                    HOLES
           1         00   00                   00   00
PHSIMP_PRO              PARTICLES                    HOLES
           1         00   00                   00   00
VACPAR_NEU              PARP PARM
                         44   42
VACPAR_PRO              PARP PARM
                         32   34
PHPARI_NEU              PARTICLES                    HOLES
           1         00   00                   00   00
PHPARI_PRO              PARTICLES                    HOLES
           1         00   00                   00   00
VACSIG_NEU         PPSP PPSM PMSP PMSM
                    22   22   21   21
VACSIG_PRO         PPSP PPSM PMSP PMSM
                    16   16   17   17
PHSIGN_NEU              PARTICLES                    HOLES
           1   00   00   00   00      00   00   00   00
PHSIGN_PRO              PARTICLES                    HOLES
           1   00   00   00   00      00   00   00   00
DIANON_NEU              PARTICLES              HOLES              TYPE
                         02                    01                 0
DIANON_PRO              PARTICLES              HOLES              TYPE
                         02                    01                 0
DIASIM_NEU              PARTICLES              HOLES              TYPE
                    02   02                01   01              0   0
DIASIM_PRO              PARTICLES              HOLES              TYPE
                    02   02                01   01              0   0
DIAPAR_NEU              PARTICLES              HOLES              TYPE
                    02   02                01   01              0   0
DIAPAR_PRO              PARTICLES              HOLES              TYPE
                    02   02                01   01              0   0
DIASIG_NEU              PARTICLES              HOLES              TYPE
              02   02   02   02      01   01   01   01    0   0   0   0
DIASIG_PRO              PARTICLES              HOLES              TYPE
              02   02   02   02      01   01   01   01    0   0   0   0
                    ----------  Numerical data  --------
MAX_MULTIP
           2      4      4
MAX_SURFAC
           0      0
MAX_MAGNET
           0      0
COULOMB
          80   79   0.25
SLOW-DOWN
         0.5  0.5
```

```
SLOW-PAIR
          0.5
EPS_HERMIT
          1.00E-14
OPTI_GAUSS
          1
GAUSHERMIT
          18    18    32
                    ---  Parameters of the HO basis  ---
BASIS_SIZE
          15   301   800.
HOMEGAZERO
          1.2
SURFAC_PAR
          86    66   1.23
SURFAC_DEF
          2     0   0.61
SURFAC_DEF
          4     0   0.10
                    ----------  Constraints  ----------
OMEGAY
          0.50
OMISOY
          0.00
OMEGA_XYZ
          0.00   0.00   0.00        0
OMEGA_RTP
          0.00   0.00   0.00        0
OMISO_RTP
          0.00   0.00   0.00        0
OMEGA_TURN
          0
MULTCONSTR
          2     0   0.01   42.   1
SURFCONSTR
          2     0   0.00   00.   0
SPINCONSTR
                     0.00    0.   0
SPICON_XYZ
                     0.00    0.   0
                     0.00    0.   0
                     0.00    0.   0
SPICON_OME
                     0.00    0.   0
                    ----  Output-file  parameters  -----
PRINT-ITER
          1     0     1
PRINT-MOME
```

```
              1      1      1
PRINT-INTR
              1
EALLMINMAX
              -12.   0.
EQUASI_MAX
              10.0
                          ------------- Files  --------------
REVIEWFILE
              HFODD.REV
REVIEW
              2
RECORDFILE
              HFODD.REC
RECORDSAVE
              1
REPLAYFILE
              HFODD.REP
WOODSAFILE
              WOODS.WFN
COULOMFILE
              HFODD.COU
COULOMSAVE
              0      0
                          ----- Starting the iteration  -----
RESTART
              0
CONT_PAIRI
              0
CONT_OMEGA
              0
READ_WOODS
              0
NILSSONPAR
              0   -1.175 -0.247   -1.175 -0.352    11.170  11.170  6.280
                          ---------- Calculate  ------------
EXECUTE
                          ---------- Terminate  ------------
ALL_DONE
```

### 5.2.6.   File tb151-a.dat

```
                          ----------  General data  ----------
NUCLIDE
              86    65
ITERATIONS
              2000
ITERAT_EPS
              0.0001
MAXANTIOSC
              5
PING-PONG
              0.01  3
CHAOTIC
              20
                          ----------  Configurations  --------
VACSIG_NEU        PPSP PPSM PMSP PMSM
                   22   22   21   21
VACSIG_PRO        PPSP PPSM PMSP PMSM
                   15   16   17   17
PHSIGN_NEU              PARTICLES                HOLES
              1   00   00   23   00        00   22   00   00
PHSIGN_PRO              PARTICLES                HOLES
              1   00   00   00   00        00   00   00   00
DIASIG_NEU              PARTICLES                HOLES              TYPE
                  00   00   00   00        00   00   00   00     0  0  0  0
DIASIG_PRO              PARTICLES                HOLES              TYPE
                  00   00   00   00        00   00   00   00     0  0  0  0
                          ----------  Numerical data  --------
SLOW-DOWN
              0.5  0.5
                          ----  Output-file  parameters  -----
PRINT-ITER
              0       0     1
                          -------------  Files  --------------
REVIEWFILE
              tb151-a.rev
COULOMFILE
              tb151.cou
COULOMSAVE
              1      1
                          -----  Starting the iteration  -----
RESTART
              0
                          ----------  Calculate  -------------

OMEGAY
```

```
           0.50
REPLAYFILE
           tb151-a-0500.rec
RECORDFILE
           tb151-a-0500.rec
EXECUTE

RESTART
           1
OMEGAY
           0.55
REPLAYFILE
           tb151-a-0500.rec
RECORDFILE
           tb151-a-0550.rec
EXECUTE

OMEGAY
           0.60
REPLAYFILE
           tb151-a-0550.rec
RECORDFILE
           tb151-a-0600.rec
EXECUTE

OMEGAY
           0.65
REPLAYFILE
           tb151-a-0600.rec
RECORDFILE
           tb151-a-0650.rec
EXECUTE

OMEGAY
           0.70
REPLAYFILE
           tb151-a-0650.rec
RECORDFILE
           tb151-a-0700.rec
EXECUTE

OMEGAY
           0.75
```

```
REPLAYFILE
          tb151-a-0700.rec
RECORDFILE
          tb151-a-0750.rec
EXECUTE

OMEGAY
          0.80
REPLAYFILE
          tb151-a-0750.rec
RECORDFILE
          tb151-a-0800.rec
EXECUTE

              ----------  Terminate  -------------
ALL_DONE
```

### 5.2.7. File `tb151-b.dat`

```
                     ----------  General data  ----------
NUCLIDE
          86    65
ITERATIONS
          2000
ITERAT_EPS
          0.0001
MAXANTIOSC
          5
PING-PONG
          0.01  3
CHAOTIC
          5
                     ----------  Configurations  --------
VACSIG_NEU      PPSP PPSM PMSP PMSM
                22   22   21   21
VACSIG_PRO      PPSP PPSM PMSP PMSM
                15   16   17   17
PHSIGN_NEU              PARTICLES              HOLES
          1   00   00   24   00      00   22   00   00
PHSIGN_PRO              PARTICLES              HOLES
          1   00   00   00   00      00   00   00   00
DIASIG_NEU              PARTICLES              HOLES          TYPE
              00   00   00   00      00   00   00   00    0  0  0  0
DIASIG_PRO              PARTICLES              HOLES          TYPE
              00   00   00   00      00   00   00   00    0  0  0  0
                     ----------  Numerical data  --------
SLOW-DOWN
          0.5  0.5
                     ----  Output-file  parameters  -----
PRINT-ITER
          0    0    1
                     -------------  Files  -------------
REVIEWFILE
          tb151-b.rev
COULOMFILE
          tb151.cou
COULOMSAVE
          1    1
                     -----  Starting the iteration  -----
RESTART
          0
                     ----------  Calculate  -------------

OMEGAY
```

```
            0.50
REPLAYFILE
            tb151-b-0500.rec
RECORDFILE
            tb151-b-0500.rec
EXECUTE

RESTART
            1
OMEGAY
            0.55
REPLAYFILE
            tb151-b-0500.rec
RECORDFILE
            tb151-b-0550.rec
EXECUTE

OMEGAY
            0.60
REPLAYFILE
            tb151-b-0550.rec
RECORDFILE
            tb151-b-0600.rec
EXECUTE

OMEGAY
            0.65
REPLAYFILE
            tb151-b-0600.rec
RECORDFILE
            tb151-b-0650.rec
EXECUTE

OMEGAY
            0.70
REPLAYFILE
            tb151-b-0650.rec
RECORDFILE
            tb151-b-0700.rec
EXECUTE

OMEGAY
            0.75
```

```
REPLAYFILE
          tb151-b-0700.rec
RECORDFILE
          tb151-b-0750.rec
EXECUTE

OMEGAY
          0.80
REPLAYFILE
          tb151-b-0750.rec
RECORDFILE
          tb151-b-0800.rec
EXECUTE

          ----------  Terminate  -------------
ALL_DONE
```

### 5.2.8.   File tb151-c.dat

```
                        ----------  General data  ----------
NUCLIDE
             86    65
ITERATIONS
             2000
ITERAT_EPS
             0.0001
MAXANTIOSC
             5
PING-PONG
             0.01  3
CHAOTIC
             5
                        ----------  Configurations  --------
VACSIG_NEU         PPSP PPSM PMSP PMSM
                    22   22   21   21
VACSIG_PRO         PPSP PPSM PMSP PMSM
                    15   16   17   17
PHSIGN_NEU              PARTICLES              HOLES
             1   00   00   23   00        00   22   00   00
PHSIGN_PRO              PARTICLES              HOLES
             1   00   00   00   00        00   00   00   00
DIASIG_NEU              PARTICLES              HOLES         TYPE
                 00   00   24   00        00   00   23   00     0  0 +1  0
DIASIG_PRO              PARTICLES              HOLES         TYPE
                 00   00   00   00        00   00   00   00     0  0  0  0
                        ----------  Numerical data  --------
SLOW-DOWN
             0.5  0.5
                        ----  Output-file  parameters  -----
PRINT-ITER
             0      0      1
                        -------------  Files  --------------
REVIEWFILE
             tb151-c.rev
COULOMFILE
             tb151.cou
COULOMSAVE
             1      1
                        -----  Starting the iteration  -----
RESTART
             0
                        ----------  Calculate  -------------

OMEGAY
```

```
            0.50
REPLAYFILE
            tb151-c-0500.rec
RECORDFILE
            tb151-c-0500.rec
EXECUTE

RESTART
            1
OMEGAY
            0.55
REPLAYFILE
            tb151-c-0500.rec
RECORDFILE
            tb151-c-0550.rec
EXECUTE

SLOW-DOWN
            0.95 0.95
OMEGAY
            0.60
REPLAYFILE
            tb151-c-0550.rec
RECORDFILE
            tb151-c-0600.rec
EXECUTE

OMEGAY
            0.65
REPLAYFILE
            tb151-c-0600.rec
RECORDFILE
            tb151-c-0650.rec
EXECUTE

OMEGAY
            0.70
REPLAYFILE
            tb151-c-0650.rec
RECORDFILE
            tb151-c-0700.rec
EXECUTE
```

```
SLOW-DOWN
          0.5  0.5
OMEGAY
          0.75
REPLAYFILE
          tb151-c-0700.rec
RECORDFILE
          tb151-c-0750.rec
EXECUTE

OMEGAY
          0.80
REPLAYFILE
          tb151-c-0750.rec
RECORDFILE
          tb151-c-0800.rec
EXECUTE

                    ----------  Terminate  -------------
ALL_DONE
```

### 5.2.9. File tb151-d.dat

```
                        ----------  General data  ----------
NUCLIDE
           86    65
ITERATIONS
           2000
ITERAT_EPS
           0.0001
MAXANTIOSC
           5
PING-PONG
           0.01  3
CHAOTIC
           5
                        ----------  Configurations  --------
VACSIG_NEU        PPSP PPSM PMSP PMSM
                  22   22   21   21
VACSIG_PRO        PPSP PPSM PMSP PMSM
                  15   16   17   17
PHSIGN_NEU                PARTICLES                HOLES
           1   00   00   23   00        00   22   00   00
PHSIGN_PRO                PARTICLES                HOLES
           1   00   00   00   00        00   00   00   00
DIASIG_NEU                PARTICLES                HOLES            TYPE
               00   00   24   00        00   00   23   00      0  0 -1  0
DIASIG_PRO                PARTICLES                HOLES            TYPE
               00   00   00   00        00   00   00   00      0  0  0  0
                        ----------  Numerical data  --------
SLOW-DOWN
           0.5  0.5
                        ----  Output-file  parameters  -----
PRINT-ITER
           0      0     1
                        -------------  Files  --------------
REVIEWFILE
           tb151-d.rev
COULOMFILE
           tb151.cou
COULOMSAVE
           1      1
                        -----  Starting the iteration  -----
RESTART
           0
                        ----------  Calculate  -------------

OMEGAY
```

```
            0.50
REPLAYFILE
            tb151-d-0500.rec
RECORDFILE
            tb151-d-0500.rec
EXECUTE

RESTART
            1
SLOW-DOWN
            0.95 0.95
OMEGAY
            0.50
REPLAYFILE
            tb151-d-0500.rec
RECORDFILE
            tb151-d-0500.rec
EXECUTE

OMEGAY
            0.55
REPLAYFILE
            tb151-d-0500.rec
RECORDFILE
            tb151-d-0550.rec
EXECUTE

OMEGAY
            0.60
REPLAYFILE
            tb151-d-0550.rec
RECORDFILE
            tb151-d-0600.rec
EXECUTE

OMEGAY
            0.65
REPLAYFILE
            tb151-d-0600.rec
RECORDFILE
            tb151-d-0650.rec
EXECUTE
```

```
OMEGAY
          0.70
REPLAYFILE
          tb151-d-0650.rec
RECORDFILE
          tb151-d-0700.rec
EXECUTE

SLOW-DOWN
          0.5  0.5
OMEGAY
          0.75
REPLAYFILE
          tb151-d-0700.rec
RECORDFILE
          tb151-d-0750.rec
EXECUTE

OMEGAY
          0.80
REPLAYFILE
          tb151-d-0750.rec
RECORDFILE
          tb151-d-0800.rec
EXECUTE

          ----------  Terminate  -------------
ALL_DONE
```

## 5.2.10.  File la132-a.dat

```
========================================================================
| This file (la132-a.dat) contains the input data for the code HFODD.  |
| It defines two consecutive runs that lead to a planar solution in 132La. |
========================================================================
                        ------------  Nuclide  ------------
NUCLIDE
            75 57
                        ----------  Convergence  -----------
ITERATIONS
            5000
ITERAT_EPS
            0.000001
MAXANTIOSC
            5
                        -------  Skyrme interaction  -------
SKYRME-STD
            1 0 0 0 0
                        ---------  Configurations  ---------
VACSIG_NEU      PPSP  PPSM  PMSP  PMSM
                19    19    19    18
VACSIG_PRO      PPSP  PPSM  PMSP  PMSM
                15    15    14    13
                        ------------  HO basis  ------------
BASIS_SIZE
            10 286 800.
SURFAC_PAR
            75 57 1.230
SURFAC_DEF
            2 0 0.000
SURFAC_DEF
            4 0 0.000
                        ----------  Output data  -----------
MAX_MAGNET
            0 3
                        -----------  File names  -----------
REVIEWFILE
            la132-a.rev
REPLAYFILE
            la132-a.rec
RECORDFILE
            la132-a.rec
                        ----------  Constraints  -----------
OMEGAY
            0.000
MULTCONSTR
            2 0 0.5  7.0    1
```

```
MULTCONSTR
          2 2 0.5  7.0    1
                      --------- Starting point ---------
RESTART
          0
                      ------------ Calculate -----------
EXECUTE
                      ----------- Symmetries -----------
SIMPLEXY
          0
SIGNATUREY
          0
PARITY
          1
TSIMPLEX3D
          0 0 0
                      --------- Configurations ---------
VACPAR_NEU          PARP  PARM
                     38    37
VACPAR_PRO          PARP  PARM
                     30    27
                      ---------- Constraints -----------
OMEGA_XYZ
          0.1443375673 0.1443375673 0.1443375673 1
OMEGA_TURN
          1
MULTCONSTR
          2 0 0.5  7.0    0
MULTCONSTR
          2 2 0.5  7.0    0
                      --------- Starting point ---------
RESTART
          1
CONT_OMEGA
          0
                      ------------ Calculate -----------
EXECUTE
                      ------------ Terminate -----------
ALL_DONE
```

### 5.2.11.   File la132-b.dat

```
=======================================================================
| This file (la132-b.dat) contains the input data for the code HFODD.  |
| It defines two consecutive runs that lead to a planar solution in 132La. |
=======================================================================
                         ------------  Nuclide  -------------
NUCLIDE
            75 57
                         ----------  Convergence  -----------
ITERATIONS
            5000
ITERAT_EPS
            0.000001
MAXANTIOSC
            5
                         -------  Skyrme interaction  -------
SKYRME-STD
            1 0 0 0 0
                         ---------  Configurations  ---------
VACSIG_NEU      PPSP  PPSM  PMSP  PMSM
                19    19    19    18
VACSIG_PRO      PPSP  PPSM  PMSP  PMSM
                15    15    14    13
                         ------------  HO basis  ------------
BASIS_SIZE
            10 286 800.
SURFAC_PAR
            75 57 1.230
SURFAC_DEF
            2 0 0.000
SURFAC_DEF
            4 0 0.000
                         ----------  Output data  -----------
MAX_MAGNET
            0 3
                         -----------  File names  -----------
REVIEWFILE
            la132-b.rev
REPLAYFILE
            la132-b.rec
RECORDFILE
            la132-b.rec
                         ----------  Constraints  -----------
OMEGAY
            0.000
MULTCONSTR
            2 0 0.5  7.0    1
```

```
MULTCONSTR
          2 2 0.5  7.0    1
                    --------- Starting point ---------
RESTART
          0
                    ------------ Calculate -----------
EXECUTE
                    ----------- Symmetries -----------
SIMPLEXY
          0
SIGNATUREY
          0
PARITY
          1
TSIMPLEX3D
          1 0 0
                    --------- Configurations ---------
VACPAR_NEU              PARP  PARM
                        38    37
VACPAR_PRO             PARP  PARM
                        30    27
                    ---------- Constraints -----------
OMEGA_XYZ
          0.0000000 0.1767766953 0.1767766953 1
OMEGA_TURN
          1
MULTCONSTR
          2 0 0.5  7.0    0
MULTCONSTR
          2 2 0.5  7.0    0
                    --------- Starting point ---------
RESTART
          1
CONT_OMEGA
          0
                    ------------ Calculate -----------
EXECUTE
                    ------------ Terminate -----------
ALL_DONE
```

### 5.2.12.  File pb208-a.dat

```
                         ----------  General data  ----------
NUCLIDE
           126   82
ITERATIONS
           500
ITERAT_EPS
           0.0000001
MAXANTIOSC
           5
PING-PONG
           0.01   3
CHAOTIC
           5
PHASESPACE
           64     64     42     42
                         ----------  Interaction  -----------
SKYRME-SET
           SLY4
SKYRME-STD
           1      0      1      1      0
                         ----------  Symmetries  ------------
ROTATION
           0
                         ----------  Configurations  --------
VACSIG_NEU      PPSP  PPSM  PMSP  PMSM
                29    29    34    34
VACSIG_PRO      PPSP  PPSM  PMSP  PMSM
                22    22    19    19
                         ----------  Numerical data  --------
OPTI_GAUSS
           0
GAUSHERMIT
           54     54     54
                         ---  Parameters of the HO basis  ---
BASIS_SIZE
           12    455    800.
SURFAC_PAR
           126    82    1.23
SURFAC_DEF
           2      0     0.0
SURFAC_DEF
           4      0     0.0
                         ----------  Constraints  -----------
OMEGAY
           0.00
MULTCONSTR
```

```
           2      0   0.10    0.0  0
                        ---- Output-file parameters -----
EALLMINMAX
           -70.  10.
                        ------------ Files -------------
REVIEWFILE
           pb208-a.rev
REVIEW
           2
RECORDFILE
           pb208-a.rec
REPLAYFILE
           pb208-a.rec
COULOMFILE
           pb208-a.cou
COULOMSAVE
           1      1
                        ---------- Calculate -------------
EXECUTE
BASIS_SIZE
           26  3654   800.
RESTART
           1
EXECUTE
                        ---------- Terminate -------------
ALL_DONE
```

### 5.2.13.  File sn120-a.dat

```
                            ----------  General data  ----------
NUCLIDE
             70    50
ITERATIONS
             500
ITERAT_EPS
             0.0000001
MAXANTIOSC
             5
PING-PONG
             0.01    3
CHAOTIC
             5
                            ----------  Interaction  -----------
SKYRME-SET
             SLY4
SKYRME-STD
             1       0       1       1       0
                            ----------   Pairing   -----------
INI_FERMI
             -8.0 -8.0
INI_DELTA
             1.0  1.0
PAIR_FORCE
             -285.88 893.375 1.0
                            ----------  Symmetries  ------------
ROTATION
             0
PAIRING
             1
HFB
             1
                            ----------  Configurations  --------
VACSIG_NEU        PPSP  PPSM  PMSP  PMSM
                  22    22    13    13
VACSIG_PRO        PPSP  PPSM  PMSP  PMSM
                  12    12    13    13
                            ---  Parameters of the HO basis  ---
BASIS_SIZE
             14    680    800.
SURFAC_PAR
             70    50    1.23
SURFAC_DEF
             2       0    0.0
SURFAC_DEF
             4       0    0.0
```

```
                          ----------  Constraints  -----------
OMEGAY
          0.00
MULTCONSTR
          2       0   0.10    0.0  0
                          ----  Output-file  parameters  -----
EALLMINMAX
          -70.  10.
                          -------------  Files  --------------
REVIEWFILE
          sn120-a.rev
REVIEW
          2
RECORDFILE
          sn120-a.rec
REPLAYFILE
          sn120-a.rec
COULOMFILE
          sn120-a.cou
COULOMSAVE
          1       1
                          ----------  Calculate  -------------
EXECUTE
                          ----------  Terminate  -------------
ALL_DONE
```

# 6.   ACKNOWLEDGMENTS

Careful reading of the manuscript by N. Schunck is greatly appreciated. This research was supported in part by the Polish Committee for Scientific Research (KBN) under Contract No. 1 P03B 059 27, by the Foundation for Polish Science (FNP), by the French-Polish integrated actions programme POLONIUM, by the U.S. Department of Energy under Contract Nos. DE-FG02-96ER40963 (University of Tennessee), and DE-AC05-00OR22725 with UT-Battelle, LLC (Oak Ridge National Laboratory), and by the computational grant from the Interdisciplinary Centre for Mathematical and Computational Modeling (ICM) of the Warsaw University.

# Subject Index

# Index of variables

# Index of input-data keywords

# Index of subroutines

# Index of files